\documentclass[12pt,fleqn]{elsarticle}

\usepackage[top=2.5cm,bottom=2.5cm,left=2.5cm,right=2.5cm]{geometry}
\usepackage{amssymb,color,ulem}
\usepackage{amsmath}
\usepackage{float,graphics}
\usepackage{natbib,euscript}
\usepackage{multirow}
\usepackage{tablefootnote}

\journal{International Journal of Heat and Mass Transfer}

\newcommand\lae[1]{\label{#1}}

\begin{document}

\begin{frontmatter}

\title{Sublimation and deposition at a solid sphere in the presence of a non-condensable gas}

\author[1]{Denize Kalempa}
\author[2]{Felix Sharipov}
\author[3]{Irina Graur}
\affiliation[1]{organization={Departamento de Ci\^encias B\'asicas e Ambientais, Escola de Engenharia de Lorena, Universidade de S\~ao Paulo},city={Lorena},postcode={12602-810}, country={Brazil}}

\affiliation[2]{organization={Departamento de F\'{\i}sica, Universidade Federal do Paran\'a, Caixa Postal 19044},city={Curitiba},postcode={81531-980}, country={Brazil}}

\affiliation[3]{organization={Aix-Marseille Universit\'e, CNRS, IUSTI UMR 7343},city={Marseille},postcode={13453}, country={France}}

\begin{abstract}
The sublimation/deposition process at a solid sphere of argon into its vapor in the presence of helium as a background gas is modeled applying the linearized Boltzmann equation, in which the McCormack model is employed for the collisional term.
The Onsager coefficients determining the mass and energy flow rates at the interface are calculated over a wide range of rarefaction parameter and for some values of molar fraction of the background gas in the mixture. Moreover, two values for the temperature of the mixture are considered on the basis of the sublimation curves of argon and krypton. To assess the influence of the interatomic interaction potential on the numerical results, the calculations are carried out for both the hard-spheres and \textit{ab-initio} potentials. The kinetic coefficients are presented as well as the flow fields around the sphere. The effect of a small temperature difference between the spherical surface and the gas mixture is analyzed. 
\end{abstract}

\begin{keyword}
Phase transition, Boltzmann equation, \textit{ab-initio} potential, mass and heat transfer.
\end{keyword}

\end{frontmatter}

\section{Introduction}

Sublimation and deposition processes are phase transitions in which a substance changes directly between solid and gas states without passing through the liquid phase. One of the very well-known examples is the sublimation of solid carbon dioxide (dry ice). When the temperature becomes higher than -78.5$^\circ$C solid carbon dioxide sublimes, turning directly from the solid phase to the gas phase. Spacecrafts and space suits can reject heat by sublimating ice into the vacuum of space \cite{CHANG2023100930}.
Freeze drying, or lyophilization, is a low temperature dehydratation process that involves freezing the product and lowering pressure, thereby removing the ice by sublimation. This technological process is used in the food and pharmaceutical industries \cite{GANGULY20121739} and the preparation of specimens for a scanning electronic microscope or a
transmission electronic microscope  \cite{Faghri:2020}.

The vapor deposition is the opposite of sublimation, i.e. it is process of vapor transforming directly to solid state without condensation. This process, sometimes called physical vapor transport (PVT), describes a variety of vacuum deposition methods which can be used to produce thin films and coatings on substrates including metals, ceramics, glass, and polymers \cite{KERN200111}. The PVT is characterized by a process in which the material undergoes sublimation at a high temperature and a low pressure  to a vapor phase, usually in a carrier gas (e.g. argon), and then deposit (condense) to a thin film or crystal \cite{Su:2020}. 
These processes are significant in various scientific and industrial applications, including vacuum system development, heat exchangers, and chemical reactors. 

In the context of kinetic theory of gases \cite{Fer02,CerB3,Son48,Sha02B}, the sublimation and deposition between two solid plates of argon in the presence of helium was studied in Ref. \cite{Sha129}. Numerous works have studied evaporation and condensation phenomena - processes analogous to sublimation and deposition - in the case of single vapor as well as in the presence of non-condensable gas confined between two parallel plates, see e.g. Refs. \cite{Pao03,Son27,Son09, Aok08,Aok15,Aok01,Son35,Sie24} and between two co-axial cylinders, see e.g. Refs. \cite{Sha28,Sha35,Sug01,Son12}.
These studies aim to understand the influence of the phase transitions on the transport phenomena in non-equilibrium gases. However, simplified models are used in the modeling, such as the hard-sphere model for the intermolecular interaction potential and hypothetical molecular masses, which may not accurately capture the complexities of real molecular interactions. Problems concerning evaporation and condensation at a spherical droplet in the presence of a non-condensable gas were studied in Refs. \cite{Margilevskii:1985,Kos11,Oni07} applying the Hamel model \cite{Ham01}. However, such a model does not provide the correct expressions of all transport coefficients. The McCormack model \cite{McC02} for the linearized Boltzmann equation allows to consider an arbitrary intermolecular interaction law and to obtain correctly the transport coefficients of the mixture. This model was used in Ref. \cite{Margilevskii:1985} to describe a droplet of water evaporating in air. A comparison of results based on hard-sphere molecular model with those based on the Lennard-Jones potential showed a significant influence of the intermolecular interaction potential on the results of the problem. Moreover, the papers \cite{Margilevskii:1985,Kos11,Oni07} showed that the mass and heat transfers due to sublimation or evaporation from a solid or liquid drop are qualitatively different from those between two parallel plates. 

In the present work, we study the mass and energy flow rates due to sublimation and deposition from a solid argon sphere surrounded by non-condensable helium. The thermal conditions, i.e. temperature, pressure, chemical compositions, considered here are the same as those in the previous work \cite{Sha129}. Therefore, the qualitatively different behavior of the spherically symmetrical flow from that of the planar one will be evident. To verify the influence of the chemical composition on the problem additional  calculations are carried out for a krypton solid sphere immersed in non-condensable helium. An specific temperature range in which the argon and krypton sublimate or deposit was chosen on the basis of the sublimation curves given in Ref. \cite{Fer06}.  
The main aim is to understand how the background or non-condensable gas, helium, influences mass and heat transfer as a result of phase transitions on the surface of the sphere and to analyze the influence of the interatomic interaction potential on the solution of the problem.

The McCormack model \cite{McC02} for the linearized Boltzmann equation is used considering both hard-sphere (HS) model and  \textit{ab-initio} (AI) potentials for the interatomic interactions. The advantage of using the AI potential is the absence of parameters extracted from experimental data.
The McCormack model is widely used in modeling of gaseous mixture flows because it satisfies the mass, momentum and energy conservation laws, H-theorem, and provides correct expressions for the  transport coefficients of the mixture such as viscosity, heat conductivity, diffusion and thermal diffusion. A comparison of the  numerical results for benchmark problems based on the linearized Boltzmann equation and direct simulation Monte Carlo method with those obtained from the McCormack model in Refs. \cite{HoM03,Sha100,Sha105} showed the reliability of the model. Moreover, a comparison between experimental data and numerical results obtained from the McCormack model for binary gas mixture flowing through long microchannels reported in Ref. \cite{Sza02} reaffirms the reliability. 

The HS model of intermolecular interaction is widely employed in modeling of rarefied gas flows. In spite of being the most simple potential, it works very well for many practical applications. However, there are situations in which a more realistic model is necessary to describe the physical phenomena properly. For instance, the thermal slip and temperature jump phenomena in gas mixtures are strongly dependent on the intermolecular interaction potential, e.g. Refs. \cite{Sha48, Sha55}. In fact, the application of the HS model leads to a qualitatively different dependence of the thermal slip and temperature jump coefficients on the mixture concentration. In the present problem the pressure and temperature jumps at the interface appears. Moreover, the results reported in Ref. \cite{Sha129} show that behaviors of same quantities, such as temperature of the mixture, change qualitatively when the HS model is replaced by AI or Lennard-Jones potentials. 

The deterministic discrete velocity method, whose details can be found in Ref. \cite{Sha02B}, is the numerical technique to solve the system of kinetic equations for the binary gas mixture under the assumption of complete phase transition for argon/krypton and diffuse scattering for helium at the sphere surface. The mass and energy fluxes at the vapor-solid interface, which are quantities of practical interest, are determined in terms of kinetic coefficients from non-equilibrium thermodynamics \cite{DeG01}.  

\section{Statement of the problem}

Let us consider a solid argon sphere of radius $R_0$ being at rest in a mixture of argon vapor and non-condensable helium. The temperature of the spherical surface is assumed to be constant and equal to $T_s$. Far from the sphere, the gas mixture is in equilibrium, characterized by temperature $T_0$, pressure $p_0$ and molar fraction $C_0$ defined as 
\begin{equation}
C_0=\frac{n_{01}}{n_0}, \quad n_0=n_{01}+n_{02},
\lae{sp1}
\end{equation}
with $n_{01}$ and $n_{02}$ denoting the number density of helium and argon/krypton, respectively. 

The rarefaction parameter, which is inversely proportional to the Knudsen number and characterizes the flow regime of the rarefied gas, is introduced as
\begin{equation}
\delta=\frac{R_0}{\ell_0},\quad \ell_0=\frac{\mu_0v_0}{p_0},
\lae{sp2}
\end{equation}
where $\ell_0$ is the equivalent molecular free path, $\mu_0$ is the viscosity of the mixture at temperature $T_0$ and $v_0$ is the characteristic molecular speed of the mixture which reads
\begin{equation}
v_0=\sqrt{\frac{2kT_0}{m}},\quad m=C_0m_1 +(1-C_0)m_2.
\lae{sp3}
\end{equation}
$m_{\alpha}$ denotes the molecular mass of species $\alpha$, $m$ is the mean molecular mass of the mixture and $k$ is the Boltzmann constant. 
It should be noted that $\ell_0$ has the order of the molecular mean free path and is more suitable to analyze the influence of the interatomic interaction potential because it does not depend on the potential cutoff distance. 

The rarefaction parameter classifies the gas flow into three regimes: free molecular ($\delta \ll 1$), transitional ($\delta \sim 1$) and viscous ($\delta \gg 1$) regime. This classification helps to select an appropriate approach to address the gas flow problem.  

The mixture is perturbed by the temperature difference $\Delta T=T_s-T_0$, which causes the pressure difference of argon/krypton $\Delta p_2=p_{2s}-p_{02}$. Here, $p_{2s}$ is the saturation pressure of argon/krypton at the temperature $T_s$ and $p_{02}$ is the partial pressure of argon/krypton in equilibrium mixture. In practice, the pressure $p_{2s}$ is related to the temperature $T_s$ by the Clapeyron equation, but formally, the differences $\Delta p_2$ and $\Delta T$ can be considered as two independent driving forces. Thus, two independent thermodynamic forces are introduced as
\begin{equation}
X_P=\frac{\Delta p_2}{p_{02}},\quad
X_T=\frac{\Delta T}{T_0},
\lae{sp4}
\end{equation}
 which are assumed to be small, i.e.
\begin{equation}
|X_P| \ll 1,\quad |X_T| \ll 1, 
\lae{sp4a}
\end{equation}
so that the kinetic equation can be linearized with respect to $X_P$ and $X_T$. 

The problem is solved in the framework of kinetic theory of gases, on the level of the velocity distribution functions $f_{\alpha}({\bf r}',{\bf v}_{\alpha})$, $\alpha$=1, 2, where ${\bf v}_{\alpha}$ is the molecular velocity of species $\alpha$ and ${\bf r}'$ is the position vector. Due to the spherical geometry of the problem, spherical coordinates $(r',\theta',\phi')$ are introduced in the physical space. The components of the molecular velocity ${\bf v}_{\alpha}$=$(v_{\alpha r},v_{\alpha \theta},v_{\alpha \phi})$ of species $\alpha$ in the mixture are written in spherical coordinates $(v_{\alpha},\theta,\phi)$ as 
\begin{equation}
\begin{split}
&v_{\alpha r}=v_{\alpha}\cos{\theta},\\
&v_{\alpha \theta}=v_{\alpha t}\cos{\phi},\\
&v_{\alpha \phi}=v_{\alpha t}\sin{\phi},
\end{split}
\lae{sp5}
\end{equation}
where
\begin{equation}
v_{\alpha t}=v_{\alpha}\sin{\theta}
\lae{sp6}
\end{equation}
is the tangential component.

The mass and energy flow rates from the sphere are defined via the distribution functions as \cite{Fer02,Cha04}
\begin{equation}
\dot{M}=4\pi R_0^2 m_2 \int v_{2r}f_2(R_0,{\bf v}_2)\, \mbox{d}{\bf v}_2, \, 
\lae{sp7}
\end{equation}
\begin{equation}
\dot{E}=2\pi R_0^2 \sum_{\alpha=1}^2 m_{\alpha}\int
v_{\alpha r}{\bf v}_{\alpha}^2
f_{\alpha}(R_0,{\bf v}_{\alpha})\, \mbox{d}{\bf v}_{\alpha}.
\lae{sp8}
\end{equation}
Note, the energy flow rate $\dot{E}$ includes heat transfer by both conductance and convection. 

We are going to calculate $\dot{M}$ and $\dot{E}$ as functions of the rarefaction parameter $\delta$, molar fraction $C_0$ and temperature $T_0$. Both $\dot{M}$ and $\dot{E}$ are proportional to $X_P$ and $X_T$ because of the assumption (\ref{sp4a}).

\section{Kinetic equation}

The distribution functions obey the system of two Boltzmann equations \cite{Fer02,Cha04}
\begin{equation}
v_{\alpha r}\frac{\partial f_{\alpha}}{\partial r'}-\frac{v_{\alpha t}}{r'}
\frac{\partial f_{\alpha}}{\partial \theta}=
\sum_{\beta=1}^2 Q(f_{\alpha},f_{\beta}),\quad \alpha=1,2,
\lae{ke01}    
\end{equation}
where $Q(f_{\alpha},f_{\beta})$ is the collision integral between species $\alpha$ and $\beta$. The assumptions of smallness of thermodynamic forces (\ref{sp4a}) allow to linearize Eq. (\ref{ke01}) by representing the distribution functions as 
\begin{equation}
f_{\alpha}(r',{\bf v}_{\alpha})=f_{0\alpha}[1+h_{\alpha}^{(P)}(r',{\bf v}_{\alpha})X_P+ h_{\alpha}^{(T)}(r',{\bf v}_{\alpha})X_T],
\lae{ke1}    
\end{equation}
where $h_{\alpha}^{(i)}$ are the perturbation functions due to thermodynamic force $X_i$ ($i$=$P, T$) and $f_{0\alpha}$ is the Maxwell distribution function in thermodynamic which reads 
\begin{equation}
f_{0\alpha}=n_{0\alpha}\left(\frac{m_{\alpha}}{2\pi kT_0}\right)^{3/2}
\exp{\left(-\frac{m_{\alpha}v_{\alpha}^2}{2kT_0}\right)}.
\lae{ke2}
\end{equation}

For convenience, dimensionless radial coordinate and molecular velocity are introduced as
\begin{equation}
r=\frac{r'}{\ell_0},\quad {\bf c}_{\alpha}=
\sqrt{\frac{m_{\alpha}}{2kT_0}}
{\bf v}_{\alpha}.
\lae{ke4}
\end{equation}
Then, the dimensionless radius of the sphere becomes the rarefaction parameter, i.e.  $r_0=R_0/\ell_0=\delta$.
Substituting the representation (\ref{ke1}) into (\ref{ke01}), and  using the dimensionless quantities  (\ref{ke4}), the linearized kinetic equations are obtained as
\begin{equation}
c_{\alpha r}\frac{\partial h_{\alpha}^{(i)}}{\partial r} - \frac{c_{\alpha t}}{r}\frac{\partial h_{\alpha}^{(i)}}{\partial \theta}=\ell_0\sqrt{\frac{m_{\alpha}}{2kT_0}}
\sum_{\beta=1}^2 \hat{L}_{\alpha \beta}h_{\alpha}^{(i)},
\quad i=P, T,
\lae{ke5}
\end{equation}
where $\hat{L}_{\alpha \beta}h_{\alpha}^{(i)}$ is the linearized collision integral. Thus, a system of two kinetic equations should be solved  for each thermodynamic force $X_P$ and $X_T$. The model proposed by McCormack \cite{McC02} for the collision integral provided in Appendix A, Eq. (\ref{ke6}), is used here. 

The dimensionless moments of the perturbation functions corresponding to partial density and temperature deviations from equilibrium are given as
\begin{equation}
\nu_{\alpha}^{(i)}(r)=\frac{n_{\alpha}-n_{0\alpha}}{n_{0\alpha}}=\frac{1}{\pi^{3/2}}\int h_{\alpha}^{(i)}
\mbox{e}^{-c_{\alpha}^2}\, \mbox{d}{\bf c}_{\alpha},
\lae{ke70}
\end{equation}

\begin{equation}
\tau_{\alpha}^{(i)}(r)=\frac{T_{\alpha}-T_0}{T_0}=\frac{2}{3\pi^{3/2}}\int \left(c_{\alpha}^2-\frac 32\right)h_{\alpha}^{(i)}
\mbox{e}^{-c_{\alpha}^2}\, \mbox{d}{\bf c}_{\alpha},
\lae{ke7}
\end{equation}
where $\mbox{d}{\bf c}_{\alpha}$=$c_{\alpha}^2\sin{\theta}\mbox{d}c_{\alpha}\mbox{d}\theta\mbox{d}\phi$.

The dimensionless radial components of the  bulk velocity and heat flux of species read
\begin{equation}
u_{\alpha}^{(i)}(r)=\frac{U_{\alpha}^{(i)}}{v_0}=
\frac{1}{\pi^{3/2}}\sqrt{\frac{m}{m_{\alpha}}}\int c_{\alpha r}h_{\alpha}^{(i)}
\mbox{e}^{-c_{\alpha}^2}\, \mbox{d}{\bf c}_{\alpha},
\lae{ke8}
\end{equation}
\begin{equation}
q_{\alpha}^{(i)}(r)=\frac{Q_{\alpha}^{(i)}}{p_0v_0}=
\frac{1}{\pi^{3/2}}\sqrt{\frac{m}{m_{\alpha}}}\int c_{\alpha r}\left(c_{\alpha}^2-\frac 52\right)h_{\alpha}^{(i)}
\mbox{e}^{-c_{\alpha}^2}\, \mbox{d}{\bf c}_{\alpha}.
\lae{ke9}
\end{equation}

Moreover, the component of the pressure tensor deviation from equilibrium reads
\begin{equation}
\Pi_{\alpha}^{(i)}(r)=\frac{P_{\alpha rr}^{(i)}-p_0}{p_0}=
\frac{1}{\pi^{3/2}}\int \left(c_{\alpha r}^2-\frac 13 c_{\alpha t}^2\right)h_{\alpha}^{(i)}
\mbox{e}^{-c_{\alpha}^2}\, \mbox{d}{\bf c}_{\alpha}.
\lae{ke10}
\end{equation}

The dimensional number densities and temperature are expressed via dimensionless quantities as 
\begin{equation}
n_{\alpha}=n_{\alpha 0}(1+\nu_{\alpha}^{(P)}X_P+\nu_{\alpha}^{(T)}X_T),
\lae{AB}
\end{equation}
\begin{equation}
T_{\alpha}=T_0(1+\tau_{\alpha}^{(P)}X_P+\tau_{\alpha}^{(T)}X_T),
\lae{AC}
\end{equation}
\begin{equation}
T=C_0T_1 + (1-C_0)T_2.
\lae{AD}
\end{equation}
In Sec. 5, the mass and energy flow rates are related to the moments $u_2^{(i)}$ and $q_{\alpha}^{(i)}$.

\section{Boundary conditions and asymptotic behaviors}

 We assume that the background gas, helium, is diffusely scattered from the spherical surface, which means that the helium atoms are fully accommodated at the surface and then reflected with the Maxwell distribution function corresponding to the temperature of the surface. Thus, from the representation (\ref{ke1}), the perturbation functions of the first species reflected from the surface due to the driven forces $X_P$ and $X_T$ are obtained as
\begin{equation}
h_1^{(P)}(r_0,{\bf c}_1)=\nu_{1w}^{(P)},\quad c_{1r}>0,
\lae{kb1a}
\end{equation}
\begin{equation}
h_1^{(T)}(r_0,{\bf c}_1)=\nu_{1w}^{(T)}+c_1^2-2, \quad c_{1r}>0.
\lae{kb1b}
\end{equation}
The constants
\begin{equation}
\nu_{1w}^{(i)}=-\frac{2}{\pi}\int_{c_{1r}<0}c_{1r}h_1^{(i)}(r_0,{\bf c}_1)\mbox{e}^{-c_1^2}\,\mbox{d}{\bf c}_1, 
\lae{kb2}
\end{equation}
are obtained from the impermeability condition at the surface.

The second species undergoes the phase transition at the boundary. The complete deposition/sublimation process is assumed at the spherical surface, which means that the incident atoms are completely absorbed by the surface and, at the same time, the surface emits atoms with the Maxwellian distribution function corresponding to the surface temperature $T_s$ and saturation pressure $p_{2s}$. Thus, the perturbation functions of the second species emmited from the surface due to the driven forces $X_P$ and $X_T$ are written as
\begin{equation}
h_2^{(P)}(r_0,{\bf c}_2)=1,\quad c_{2r}>0,
\lae{kb4a}
\end{equation}
\begin{equation}
h_2^{(T)}(r_0,{\bf c}_2)=c_2^2-\frac 52,\quad c_{2r}>0.
\lae{kb4b}
\end{equation}

Far from the sphere, i.e. at $r\rightarrow \infty$, the perturbation functions of both species $\alpha$ vanishes so that 
\begin{equation}
h_{\alpha \infty}^{(i)}=\lim_{r\rightarrow \infty}h_{\alpha}^{(i)}(r,{\bf c})=0, \quad i=P, T.
\lae{kb6}
\end{equation}

\section{Thermodynamic analysis}

According to the thermodynamics of irreversible processes \cite{DeG01}, thermodynamic fluxes $J_P$ and $J_T$ corresponding to the thermodynamic forces $X_P$ and $X_T$ can be expressed via a matrix of kinetic coefficients $\Lambda_{ij}$ ($i, j$=$P,T$) as
\begin{equation}
\begin{split}
&J_P=\Lambda_{\mbox{\tiny{PP}}}X_P + \Lambda_{\mbox{\tiny{PT}}}X_T,\\
&J_T=\Lambda_{\mbox{\tiny{TP}}}X_P + \Lambda_{\mbox{\tiny{TT}}}X_T.
\end{split}
\lae{te1}
\end{equation}
If we choose the fluxes $J_P$ and $J_T$ so that the entropy production $\sigma$ is expressed as
\begin{equation}
\sigma=J_PX_P +J_TX_T
\lae{AJ}
\end{equation}
then, the matrix $\Lambda_{ij}$ will be symmetric, i.e. it obeys the reciprocal relation. The symmetry reduces the number of independent coefficients determining the solution. Its fulfillment is used to control numerical error. However, the flow rates $\dot{M}$ and $\dot{E}$ do not satisfy the condition (\ref{AJ}), hence it is convenient to expresse them via some fluxes satisfying (\ref{AJ}). 

A general approach to reciprocal relations in gaseous mixtures based on the Boltzmann equation elaborated in Refs. \cite{Sha62,Sha81}  states that $J_P$ is proportional to the mean bulk velocity of the mixture, while $J_T$ is proportional to the heat flux through the mixture relatively to its mean bulk velocity. Following these results, the dimensionless fluxes are introduced as 
\begin{equation}
J_P=(1-C_0)u_2(\delta),
\lae{te8}
\end{equation}
\begin{equation}
J_T=C_0q_1(\delta) + (1-C_0)q_2(\delta),
\lae{te9}
\end{equation}
where the dimensionless bulk velocity $u_2$, and heat fluxes $q_1$ and $q_2$, are calculated via the perturbation functions. Note that the representation (\ref{ke1}) allows to write these quantities as
\begin{equation}
u_2=u_2^{(P)}X_P + u_2^{(T)}X_T
\lae{te10}
\end{equation}
\begin{equation}
q_{\alpha}=q_{\alpha}^{(P)}X_P + q_{\alpha}^{(T)}X_T, \quad \alpha=1, 2,
\lae{te11}
\end{equation}
where $u_2^{(i)}$ and $q_{\alpha}^{(i)}$ ($i=P, T$) are given in (\ref{ke8}) and (\ref{ke9}). As a consequence, the dimensionless kinetic coefficients are obtained as 
\begin{equation}
\Lambda_{\mbox{\tiny{PP}}}=(1-C_0)u_2^{(P)},
\lae{t12}
\end{equation}
\begin{equation}
\Lambda_{\mbox{\tiny{PT}}}=(1-C_0)u_2^{(T)},
\lae{te12}
\end{equation}
\begin{equation}
\Lambda_{\mbox{\tiny{TP}}}=C_0q_1^{(P)}+(1-C_0)q_2^{(P)},
\lae{te13}
\end{equation}
\begin{equation}
\Lambda_{\mbox{\tiny{TT}}}=C_0q_1^{(T)}+(1-C_0)q_2^{(T)}.
\lae{te14}
\end{equation}
Since the matrix of kinetic coefficients is symmetric, the reciprocity relation $\Lambda_{\mbox{\tiny{PT}}}$=$\Lambda_{\mbox{\tiny{TP}}}$ holds. Moreover, it is worth mentioning that the matrix of kinetic coefficients must be positive definite to keep the entropy production positive. 
The physical meaning of each kinetic coefficient is as follows: $\Lambda_{\mbox{\tiny{PP}}}$ describes the mass flow of argon/krypton caused by the driven force $X_P$; $\Lambda_{\mbox{\tiny{TT}}}$ is the ordinary heat flux through the mixture caused by the driven force $X_T$;  $\Lambda_{\mbox{\tiny{PT}}}$ is the mass flow of argon/krypton due to $X_T$; and $\Lambda_{\mbox{\tiny{TP}}}$ is the heat flux of the mixture due to $X_P$.  

Once the kinetic coefficients are known, the fluxes $J_P$ and $J_T$ are calculated by (\ref{te1}). The relations of the mass $\dot{M}$ and energy $\dot{E}$ flow rates to the thermodynamic fluxes are obtained from (\ref{sp7}), (\ref{sp8}), (\ref{te8}) and (\ref{te9}) as
\begin{equation}
\dot{M}=4\pi R_0^2n_0v_0m_2J_P,
\lae{te15}
\end{equation}
\begin{equation}
\dot{E}=4\pi R_0^2v_0p_0\left(J_T+\frac 52 J_P\right).
\lae{te16}    
\end{equation}
Since $\Lambda_{\mbox{\tiny{PT}}}=\Lambda_{\mbox{\tiny{TP}}}$, the quantities given in (\ref{te15}) and (\ref{te16}) are determined by three independent kinetic coefficients only. In turn, these coefficients depend on the rarefaction parameter $\delta$ and equilibrium molar fraction $C_0$.


\section{Free molecular regime}

In the free molecular regime, $\delta \ll 1$, the collision term in the Boltzmann equation can be neglected and the problem can be solved analytically. Thus, the integral-differential equation (\ref{ke5}) is reduced to the differential equation
\begin{equation}
c_{\alpha r}\frac{\partial h_{\alpha}^{(i)}}{\partial r}
-\frac{c_{\alpha t}}{r} \frac{\partial h_{\alpha}^{(i)}}{\partial \theta}=0,\quad i=P, T,
\lae{fm1}
\end{equation}
whose solution must satisfy the boundary conditions and asymptotic behaviors given in Sec. 4. It is worth noting that in the free molecular regime the velocity distribution functions of both species incident to the surface are not perturbed by molecular collisions. This means that $h_1^{(i)}$=0 in (\ref{kb2}).  

The method of the characteristics is used to solve the differential equations (\ref{fm1}) subject to the conditions given in Sec. 4 for each species. Thus, the solutions for the  thermodynamic force $X_P$ are given as  
\begin{equation}
h_1^{(P)}(r,{\bf c}_1)=0,\quad 0 \le \theta \le \pi,
\lae{fm2}
\end{equation}

\begin{equation}
h_2^{(P)}(r,{\bf c}_2)=
\begin{cases}
1,\quad 0 \le \theta \le \theta_0,\\
0,\quad \theta_0 < \theta \le \pi,
\end{cases}
\lae{fm2a}
\end{equation}
where 
\begin{equation}
\theta_0=\arcsin{\left(\frac{r_0}{r}\right)}.   
\lae{fm3}
\end{equation}

Similarly, the solutions for the thermodynamic force $X_T$ read
\begin{equation}
h_1^{(T)}(r,{\bf c}_1)=
\begin{cases}
\begin{split}
c_1^2-2,\quad 0\le \theta \le \theta_0,\\
0,\quad \theta_0 < \theta \le \pi.
\end{split}
\end{cases}
\lae{fm40}
\end{equation}
\begin{equation}
h_2^{(T)}(r,{\bf c}_2)=
\begin{cases}
\begin{split}
c_2^2-\frac 52,\quad 0\le \theta \le \theta_0,\\
0,\quad \theta_0 < \theta \le \pi.
\end{split}
\end{cases}
\lae{fm4}
\end{equation}

Substituting (\ref{fm2}) and (\ref{fm2a}) into (\ref{ke70})-(\ref{ke10}), we obtain that the moments due to $X_P$ for the first species are zero, while those for the second one read
\begin{equation}
\nu_2^{(P)}(r)=-\frac 12-\frac{\sqrt{1-(r_0/r)^2}}{2},
\lae{fm6}
\end{equation}
\begin{equation}
\tau_2^{(P)}(r)=0,
\lae{fm6a}
\end{equation}
\begin{equation}
u_2^{(P)}(r)=\frac{1}{2\sqrt{\pi}}\sqrt{\frac{m}{m_2}}\left(\frac{r_0}{r}\right)^2,
\lae{fm7}
\end{equation}
\begin{equation}
q_2^{(P)}(r)=-\frac{1}{4\sqrt{\pi}}
\sqrt{\frac{m}{m_2}}\left(\frac{r_0}{r}\right)^2,
\lae{fm7a}
\end{equation}
\begin{equation}
\Pi_2^{(P)}(r)=
\frac{\sqrt{r^2-r_0^2}}{4}\left(\frac{r_0}{r}\right)^2.
\lae{fm8}
\end{equation}

Similarly, substituting (\ref{fm40}) and (\ref{fm4}) into (\ref{ke70})-(\ref{ke10}), we obtain the moments due to $X_T$ as
\begin{equation}
\nu_1^{(T)}(r)=-\frac 14 +\frac{\sqrt{1-(r_0/r)^2}}{4},
\lae{fm90}
\end{equation}
\begin{equation}
\tau_1^{(T)}(r)=\frac 12 -\frac{\sqrt{1-(r_0/r)^2}}{2},
\lae{fm100}
\end{equation}
\begin{equation}
u_1^{(T)}(r)=0,
\lae{fm110}
\end{equation}
\begin{equation}
q_1^{(T)}(r)=\frac{1}{\sqrt{\pi}}\sqrt{\frac{m}{m_1}}
\left(\frac{r_0}{r}\right)^2,
\lae{fm120}
\end{equation}
\begin{equation}
\Pi_1^{(T)}(r)=\frac 18 \left(\frac{r_0}{r}\right)^2
\sqrt{1-(r_0/r)^2},
\lae{fm130}
\end{equation}
\begin{equation}
\nu_2^{(T)}(r)=2\nu_1^{(T)}(r),
\lae{fm9}
\end{equation}
\begin{equation}
\tau_2^{(T)}(r)=\tau_1^{(T)}(r),
\lae{fm10}
\end{equation}
\begin{equation}
u_2^{(T)}(r)=-\frac{1}{4\sqrt{\pi}}\sqrt{\frac{m}{m_2}}\left(
\frac{r_0}{r}\right)^2,
\lae{fm11}
\end{equation}
\begin{equation}
q_2^{(T)}(r)=\frac{9}{8\sqrt{\pi}}\sqrt{\frac{m}{m_2}}
\left(\frac{r_0}{r}\right)^2,
\lae{fm12}
\end{equation}
\begin{equation}
\Pi_2^{(T)}(r)=0.
\lae{fm13}
\end{equation}
Substituting the moments $u_2^{(i)}$, $q_1^{(i)}$ and $q_2^{(i)}$ ($i$=$P, T$) into Eqs. (\ref{t12})-(\ref{te14}) the kinetic coefficients are obtained as 
\begin{equation}
\Lambda_{\mbox{\tiny{PP}}}=\frac{1}{2\sqrt{\pi}}\sqrt{\frac{m}{m_2}}(1-C_0),
\lae{fm14}
\end{equation}
\begin{equation}
\Lambda_{\mbox{\tiny{PT}}}=\Lambda_{\mbox{\tiny{TP}}}=-\frac{1}{4\sqrt{\pi}}\sqrt{\frac{m}{m_2}}(1-C_0),
\lae{fm15}
\end{equation}
\begin{equation}
\Lambda_{\mbox{\tiny{TT}}}=\frac{1}{\sqrt{\pi}}\left[
C_0\sqrt{\frac{m}{m_1}} + \frac 98(1-C_0)\sqrt{\frac{m}{m_2}}
\right].
\lae{fm16}
\end{equation}

Thus, from (\ref{te15}) and (\ref{te16}), the linear relations (\ref{te1}) and the kinetic coefficients (\ref{fm14})-(\ref{fm16}), the mass and energy flow rates are obtained as
\begin{equation}
\dot{M}=2\sqrt{\pi}R_0^2m_2n_{20}\sqrt{\frac{2kT_0}{m_2}}\left(X_P-\frac 12 X_T\right),
\lae{fm17}
\end{equation}

\begin{equation}
\dot{E}=
4\sqrt{\pi}R_0^2p_{20}\sqrt{\frac{2kT_0}{m_2}}\left[X_P+\left(\frac 12+\frac{C_0}{(1-C_0)}\sqrt{\frac{m_2}{m_1}}\right) X_T\right].
\lae{fm18}
\end{equation}

\section{Transitional regime}

To verify the influence of the temperature on the solution of the problem, two reference temperatures are used, namely $T_0$ = 50 K and 70 K. These temperature values were chosen on the basis of the sublimation curve of argon and krypton provided in Ref. \cite{Fer06}. The calculations for HS potential were carried out only for the mixture helium-argon. The viscosities of helium and argon calculated in Ref. \cite{Sha129} are used here. Thus, using this data for the temperatures $T_0$=50 K and 70 K, the atomic diameter ratio $d_2/d_1$ for the mixture helium-argon was calculated as 2.109 and 2.031, respectively.

To consider an arbitrary gas rarefaction, the system of kinetic equations (\ref{ke5}) subject to the boundary conditions given in Sec. 4 must be solved numerically for each thermodynamic force. However, a peculiarity inherent to gas flows around a convex body is the discontinuity of the distribution function at the spherical surface \cite{Son13}, which must be treated carefully when a finite difference scheme is used. To overcome this difficulty, the method proposed in Ref. \cite{Nar02} is employed here. The method consists in the split of the solution into two parts as
\begin{equation}
h_{\alpha}^{(i)}(r,{\bf c}_{\alpha})=
h_{\alpha 0}^{(i)}(r,{\bf c}_{\alpha})+\tilde{h}_{\alpha}^{(i)}(r,{\bf c}_{\alpha}),
\lae{m1}
\end{equation}
where the function $h_{\alpha 0}^{(i)}$ satisfies the following differential equation
\begin{equation}
c_{\alpha r}\frac{\partial h_{\alpha 0}^{(i)}}{\partial r}-\frac{c_{\alpha t}}{r}\frac{\partial h_{\alpha 0}^{(i)}}{\partial \theta}+\delta_{\alpha}h_{\alpha 0}^{(i)}
=0,\quad i=P, T,
\lae{m2}
\end{equation}
with  
\begin{equation}
\delta_{\alpha}=\ell_0\sqrt{\frac{m_{\alpha}}{2kT_0}}\gamma_{\alpha}.
\lae{m20}
\end{equation}
The quantity $\gamma_{\alpha}$ is given in Appendix A, Eq. (\ref{ke19}).

The boundary conditions for solving (\ref{m2}) for each thermodynamic force are the same as those used in the free molecular regime; see Eqs. (\ref{fm2}), (\ref{fm2a}), (\ref{fm40}) and (\ref{fm4}).

The function $\tilde{h}_{\alpha}^{(i)}$ satisfies Eq. (\ref{ke5}) just changing $h_{\alpha}^{(i)}$ by $\tilde{h}_{\alpha}^{(i)}$. However, the boundary conditions obtained from (\ref{kb1a}), (\ref{kb1b}), (\ref{kb4a}) and (\ref{kb4b}) read
\begin{equation}
\tilde{h}_1^{(i)}(\delta,{\bf c}_1)=\tilde{\nu}_{1 w}^{(i)}, \quad c_{1 r}> 0,
\lae{m2a}
\end{equation}
\begin{equation}
\tilde{h}_2^{(i)}(\delta,{\bf c}_2)=0, \quad c_{2 r}> 0,
\lae{m2a1}
\end{equation}
where
\begin{equation}
\tilde{\nu}_{1 w}^{(i)}=-\frac{2}{\pi}\int_{c_{1r}<0}c_{1r}\tilde{h}_1^{(i)}(\delta,{\bf c}_1)\mbox{e}^{-c_1^2}\,\mbox{d}{\bf c}_1. 
\lae{m2b}
\end{equation}

The method of characteristics applied to solve Eq. (\ref{m2}) results the solutions  
\begin{equation}
h_{10}^{(P)}(r,{\bf c}_1)=0,\quad 0 \le \theta \le \pi,
\lae{m3}
\end{equation}
\begin{equation}
h_{20}^{(P)}(r,{\bf c}_2)=
\begin{cases}
\mbox{e}^{-\delta_2S/c_2},\quad 0\le \theta \le \theta_0,\\
0,\quad 0< \theta \le \pi,
\end{cases}
\lae{m4}
\end{equation}
\begin{equation}
h_{1 0}^{(T)}(r,{\bf c}_1)=
\begin{cases}
\begin{split}
\left(c_1^2-2\right)
\mbox{e}^{-\delta_1S/c_1},\quad 0\le \theta \le \theta_0,\\
0,\quad \theta_0 < \theta \le \pi,
\end{split}
\end{cases}
\lae{m5}
\end{equation}

\begin{equation}
h_{2 0}^{(T)}(r,{\bf c}_2)=
\begin{cases}
\begin{split}
\left(c_2^2-\frac 52\right)
\mbox{e}^{-\delta_2S/c_2},\quad 0\le \theta \le \theta_0,\\
0,\quad \theta_0 < \theta \le \pi.
\end{split}
\end{cases}
\lae{m5a}
\end{equation}
The angle $\theta_0$ is given in (\ref{fm3}) and $S$ is the distance along the characteristic line given as 
\begin{equation}
S=r\cos{\theta}-\sqrt{r_0^2-r^2\sin{\theta}}.
\lae{m6}
\end{equation}

The representation (\ref{m1}) also implies the split of the moments given in Eqs.  (\ref{ke70})-(\ref{ke10}) into two parts as
\begin{equation}
\nu_{\alpha}^{(i)}(r)=\nu_{\alpha 0}^{(i)}(r) + \tilde{\nu}_{\alpha}^{(i)}(r),
\lae{m7}
\end{equation}
\begin{equation}
\tau_{\alpha}^{(i)}(r)=\tau_{\alpha 0}^{(i)}(r) + \tilde{\tau}_{\alpha}^{(i)}(r),
\lae{m8}
\end{equation}
\begin{equation}
u_{\alpha}^{(i)}(r)=u_{\alpha 0}^{(i)}(r) + \tilde{u}_{\alpha}^{(i)}(r),
\lae{m9}
\end{equation}
\begin{equation}
q_{\alpha}^{(i)}(r)=q_{\alpha 0}^{(i)}(r) + \tilde{q}_{\alpha}^{(i)}(r),
\lae{m10}
\end{equation}
\begin{equation}
\Pi_{\alpha}^{(i)}(r)=\Pi_{\alpha 0}^{(i)}(r) + \tilde{\Pi}_{\alpha}^{(i)}(r).
\lae{m11}
\end{equation}
The solutions given in Eqs. (\ref{m3})-(\ref{m5a}) allow to obtain the moments with subscript $0$ due to the thermodynamic forces $X_P$ and $X_T$. These moments are given as Supplementary Material.

The system of kinetic equations for the functions $\tilde{h}_{\alpha}^{(P)}$ and $\tilde{h}_{\alpha}^{(T)}$ subject to the corresponding boundary conditions and asymptotic behaviors was solved numerically by using the discrete velocity method with numerical error of 0.1\% for the kinetic coefficients. The Gaussian quadrature was used to discretize  the molecular velocity and to calculate the moments of the perturbation functions. The numerical values of nodes and weights as well as the technique to calculate them can be found in Ref. \cite{Kry02}. The derivatives which appear in the kinetic equations were approximated by a central finite difference scheme. The accuracy of the calculations was estimated by varying the grid parameters $N_r$, $N_{\theta}$ and $N_c$ corresponding to the number of nodes in the radial coordinate $r$, angle $\theta$ and molecular speed $c_{\alpha}$. Moreover, the maximum value of the radial distance from the sphere which defines the gas flow domain, denoted by $r_{max}$, was varied. The reciprocity relation $\Lambda_{PT}=\Lambda_{TP}$ was verified numerically as an additional criterion of the numerical accuracy. First, the numerical calculations were carried out for the mixture helium-argon with mass ratio $m_2/m_1$=9.981 by considering HS and AI potentials of interatomic interaction. Two values of temperature were considered, namely $T_0$=50 K and 70 K. The molar fraction of helium was set as $C_0$= 0,1, 0,5 and 0.9. Some values of rarefaction parameter were considered, specifically $\delta$=0.01, 0.1, 0,5, 1, 2, 5 and 10, to cover the free molecular, transition and continuum flow regimes. The parameters $N_c$ and $N_{\theta}$ were set as 12 and 400, respectively, while the distance $r_{max}$ was set as $\delta$ varied so that the increment in the radial distance $\Delta r \sim$ $10^{-3}$. For instance, for $\delta$=1, the distance $r_{max}$ was set as 40 and $N_r$=40000. Second, to verify the influence of the mass ratio on the solution of the problem, the calculations were carried out for the mixture helium-krypton, with $m_2/m_1$=20.94, at $T_0$=70 K by considering AI potential. Once again the molar fraction of helium was chosen as $C_0$=0.1, 0.5 and 0.9, but only three values of rarefaction parameter were considered, namely $\delta$=0,1, 1 and 10. The kinetic coefficients and the macroscopic characteristics are presented below.

\section{Results}

The results for the kinetic coefficients, as functions of the rarefaction parameter $\delta$ and molar fraction $C_0$ for both HS and AI potentials are presented in Tables \ref{tabA} and \ref{tabD} at $T_0$=50 K and 70 K, respectively. 

It is worth noting that  the kinetic coefficients $\Lambda_{\mbox{\tiny{PP}}}$ and $\Lambda_{\mbox{\tiny{TT}}}$ are always positive, while the coefficient $\Lambda_{\mbox{\tiny{PT}}}$ can be positive or negative  depending on the conditions. Moreover, $\Lambda_{\mbox{\tiny{TP}}}$=$\Lambda_{\mbox{\tiny{PT}}}$ and the values of $\Lambda_{\mbox{\tiny{PT}}}$ are quite smaller than those of $\Lambda_{\mbox{\tiny{PP}}}$ and $\Lambda_{\mbox{\tiny{TT}}}$, the matrix of kinetic coefficients is positive definite, as expected from thermodynamic of irreversible processes \cite{DeG01}.  

\subsection{Influence of the gas rarefaction}

According to Tables \ref{tabA} and \ref{tabD}, for  fixed values of the molar fraction $C_0$, the kinetic coefficients vary significantly with increasing of the rarefaction parameter. For instance, in case of HS potential, at $T_0=50$ K, $C_0$=0.1 and $\delta$=10, the maximum relative deviations of the kinetic coefficients $\Lambda_{\mbox{\tiny{PP}}}$, $\Lambda_{\mbox{\tiny{PT}}}$ and $\Lambda_{\mbox{\tiny{TT}}}$ from their corresponding free molecular values are around 32\%, 84\% and 73\%, respectively, see Table \ref{tabA}. When the molar fraction increases from 0.1 to 0.5, (with $\delta$=10 and $T_0$=50 K), Table \ref{tabA} shows that the maximum relative deviation of $\Lambda_{PT}$ from its value in free molecular regime is larger than 100\%. In general, the kinetic coefficient $\Lambda_{\mbox{\tiny{TT}}}$ always decreases as the rarefaction parameter increases, while the kinetic coefficients $\Lambda_{\mbox{\tiny{PP}}}$ and $\Lambda_{\mbox{\tiny{PT}}}$ can either increase or decrease by increasing the rarefaction parameter. Notably, even the sign of $\Lambda_{\mbox{\tiny{PT}}}$ can change as the rarefaction parameter increases. 

In case of a planar flow, and at arbitrary molar fraction $C_0$, the coefficient $\Lambda{\mbox{\tiny{PP}}}$ always decreases monotonically as the rarefaction parameter increases, see e.g. Ref. \cite{Sha129}. 
Here, the qualitative behavior of this coefficient is different for each value of molar fraction $C_0$. From the data presented in Table \ref{tabA}, at $T_0$=50 K, $C_0$=0.1 and \textit{ab-initio} potential, $\Lambda_{\mbox{\tiny{PP}}}$ increases when $\delta$ varies from 0 to 10. In the 
limit of a single vapor this behavior is in agreement to that given in Ref. \cite{Che06}. However, at $C_0$=0.5 the coefficient   $\Lambda_{\mbox{\tiny{PP}}}$ increases up to a maximum value and then decreases, while at $C_0$=0.9 the coefficient decreases when $\delta$ varies from 0 to 10. Thus, the dependence of $\Lambda_{\mbox{\tiny{PP}}}$ on $\delta$ strongly depends on the value of molar fraction $C_0$. For a better visualization of such a profile, it is provided as Supplementary Material. 

\subsection{Influence of background gas}

The influence of the background gas, helium, on the kinetic coefficients can be analyzed from the tabulated data for fixed values of rarefaction parameter and different values of molar fraction $C_0$. As expected, the coefficient $\Lambda_{\mbox{\tiny{PP}}}$ always decreases as $C_0$ increases. For instance, according to Table \ref{tabA}, for both interaction potentials and at $\delta$=1, the coefficient $\Lambda_{\mbox{\tiny{PP}}}$ decreases by approximately 60\% when $C_0$ increases from 0.1 to 0.5. The coefficient $\Lambda_{\mbox{\tiny{PT}}}$, which can be positive or negative depending on $C_0$, also decreases monotonically by increasing the molar fraction. For example, in case of HS interaction and $\delta$=1, the coefficient $\Lambda_{\mbox{\tiny{PT}}}$ decreases by nearly 100\% when $C_0$ increases from 0.1 to 0.9. In contrast, the coefficient $\Lambda_{\mbox{\tiny{TT}}}$ increases when $C_0$ changes from 0.1 to 0.5, but decreases when $C_0$ changes from 0.5 to 0.9. These observations highlight the significant role playing by the background gas in the sublimation and deposition processes as it significantly affects both the mass, see Eq. (\ref{te15}), and energy, see Eq. (\ref{te16}), flow rates via $J_P$ and $J_T$, defined in Eq. (\ref{te1}).

\subsection{Influence of the interatomic potential}

In the free molecular regime ($\delta \rightarrow 0$), the kinetic coefficients given by Eqs. (\ref{fm14})-(\ref{fm16}) and are independent of the interatomic potential.
However, in the transitional and continuum regimes, the tabulated results show a  strong dependence of the kinetic coefficients on the interatomic potential. According to Tables \ref{tabA} and \ref{tabD}, the maximum relative difference in the coefficient $\Lambda_{PP}$ between the HS and AI potentials occurs at $C_0$=0.9 and $\delta$=10, and reaches 18\% at $T_0$=50 K and 10\% at $T_0$=70 K. Similarly,  for $\Lambda_{TT}$, the maximum relative difference is around 15\%  at both temperatures for $C_0$=0.5 and $\delta$=10. The coefficient $\Lambda_{PT}$, however, is much more sensitive to the interatomic potential. For instance,  for $C_0$=0.1 and $\delta$=10, the maximum relative difference for $\Lambda_{PT}$ is around 48\% at $T_0$=50 K and 41\% at $T_0$=70 K. The difference decreases by decreasing $\delta$. More notably,  for $C_0$=0.5 and $\delta$=10, the maximum relative difference for $\Lambda_{PT}$ exceeds 100\%. In this case, the use of the AI potential leads to  a sign change of the coefficient $\Lambda_{PT}$. Thus, the strong dependence of cross-coupling effects on the interatomic potential is observed. Therefore, the HS model is not appropriate to describe gas flows with phase transition.     

\subsection{Influence of the temperature}

The influence of the reference temperature $T_0$ on the kinetic coefficients is negligible when $\delta < 0.1$ and vanishes in the free molecular regime. As $\delta$ increases, the kinetic coefficients depend on the temperature $T_0$, although the effect remains small for the coefficients $\Lambda_{\mbox{\tiny{PP}}}$ and $\Lambda_{\mbox{\tiny{TT}}}$.  For example, for the HS potential, the maximum relative difference between the results for the coefficients $\Lambda_{\mbox{\tiny{PP}}}$ and $\Lambda_{\mbox{\tiny{TT}}}$ at $T_0$=50 K and 70 K is less than 3\% and 1\%, respectively. Similarly, when using the AI potential, the maximum relative difference for the same coefficients is below 4\% and 2\%, respectively. In contrast, the coefficient $\Lambda_{\mbox{\tiny{PT}}}$ is significantly more sensitive to temperature variations. According to the tabulated data, this effect can be substantial in some cases. For instance, in case of AI potential and at $C_0$=0.5 and $\delta$=10, the relative difference between the results obtained for $\Lambda_{\mbox{\tiny{PT}}}$ coefficient by considering 50K and 70 K is approximately 93\%. These results emphasize that while the temperature has a minor effect on the diagonal transport coefficients, it can have a pronounced impact on the cross-coupling term $\Lambda_{\mbox{\tiny{PT}}}$, particularly in the continuum regime.

\subsection{Influence of sublimating species}

 The kinetic coefficients obtained for the helium-krypton mixture are provided in Table \ref{tabFkr} as functions of the molar fraction, $C_0$, and rarefaction parameter, $\delta$, of the mixture. The comparison between the helium-argon and helium-krypton  results reveals that the kinetic coefficients are very sensitive to the mass ratio $m_2/m_1$. The maximum relative difference in the coefficients $\Lambda_{\mbox{\tiny{PP}}}$, $\Lambda_{\mbox{\tiny{PT}}}$ and $\Lambda_{\mbox{\tiny{TT}}}$ between the two mixtures is around 18\%, 35\% and 43\%, respectively, at $C_0$=0.5 and $\delta$=10. Among them, the most sensitive coefficient to the mass ratio is $\Lambda_{\mbox{\tiny{TT}}}$, with the smallest observed relative difference still reaching about 10\% at $C_0$=0.1 and $\delta$=0.1. These results highlight the significant role played by the mass ratio in determining the transport properties of binary gas mixtures, particularly in the transitional and continuum regimes.

\subsection{Examples of $J_P$ and $J_T$}\label{sec:JP_JT}

Tables \ref{tabJP1} and \ref{tabJP2} provide the values of $J_P$ and $J_T$ obtained by assuming the temperature difference between the sphere temperature and that of the mixture in equilibrium  equal to $\Delta T$=$T_s-T_0$=0.1 K. This temperature difference leads to thermodynamic forces $X_T$=0.002 at $T_0$=50 K and $X_T$=0.0014 at $T_0$=70 K. The thermodynamic force $X_P$ is calculated via the saturation pressure expression given in Ref. \cite{Fer06} and provided as Supplementary Material. Thus, the values of $X_P$ obtained at $T_0$=50 K and 70 K are equal to 0.042 and 0.025, respectively. Since $\Delta T >0$, $J_P$ is always positive so that a net mass flux is directed from the sphere toward the gaseous phase. The heat flux can be either positive or negative, which means that the heat can flow either away or towards the spherical surface. However, as indicated by Eq. (\ref{te16}), the energy flux is always positive, i.e. it is directed away from the spherical surface, from the hotter to the colder region in the gas medium. 

In Table \ref{tabdim}, the dimension mass and energy flow rates, given in (\ref{te15}) and (\ref{te16}), at $T_0$=50 K are presented for the same values of molar fraction $C_0$ considered in the previous tables and for three values of rarefaction parameter, $\delta$=0.1, 1 and 10. These quantities depend on the radius $R_0$ of the sphere, which is calculated via rarefaction parameter $\delta$ and equivalent mean free path $\ell_0$, as given in (\ref{sp2}). The viscosity of the gas mixture appearing in $\ell_0$ depends on the temperature $T_0$ and molar fraction $C_0$, and the values provided in Ref. \cite{Sha126} are used here. According to Table \ref{tabdim}, for fixed $\delta$, the larger $C_0$ the smaller the mass flow rate $\dot{M}$. Moreover,the energy flow rate is positive and also decreases by increasing the molar fraction $C_0$. Under the conditions assumed in the calculations, the radius of the sphere, calculated from (\ref{sp2}), changes with $C_0$ due to the dependence of the equivalent molecular mean free path on the viscosity and mean molecular mass of the mixture.      

\subsection{Flow fields}

The profiles of the density and temperature deviations from their equilibrium values for each species of the mixture helium-argon due to the thermodynamic force $X_P$, i.e. $\nu_{\alpha}^{(P)}$ and $\tau_{\alpha}^{(P)}$ ($\alpha$=1, 2),  as functions of the distance $r/\delta$ are shown in Figures \ref{fig1}-\ref{fig3} for molar fraction $C_0$=0.1, 0.5 and 0.9. The reference temperature is 50 K and three values of rarefaction parameter were considered, namely $\delta$=0.1, 1 and 10. The dimensionless radial distance $r/\delta$=$r'/R_0$ is introduced to facilitate the comparison of results for different values of $\delta$. Similarly, the profiles of the density and temperature deviations of each species due to the thermodynamic force $X_T$, i.e. $\nu_{\alpha}^{(T)}$ and $\tau_{\alpha}^{(T)}$, as functions of distance $r/\delta$, are shown in Figures \ref{fig4}-\ref{fig6} at the same reference temperature and for the same values of molar fraction and rarefaction parameter. The dependence of these macroscopic characteristics on the interatomic interaction potential is negligible, so that only the profiles corresponding to AI potential are shown in the aforementioned figures. 

As one can see from these figures, the density and temperature deviations of species significantly depend on the rarefaction parameter and molar fraction of the background gas. According to Figures \ref{fig1}-\ref{fig3}, $\nu_1^{^(P)}$ always decreases in the Knudsen layer, while $\nu_2^{(P)}$ always increases. As expected, the larger the molar fraction $C_0$ the larger such an effect. The temperature deviations $\tau_1^{(P)}$ and $\tau_2^{(P)}$ are always negative in the Knudsen layer and their magnitudes decrease with increasing $C_0$. From Figures \ref{fig4}-\ref{fig6} one can see that the density deviations $\nu_1^{(T)}$ and $\nu_2^{(T)}$ are always negative in the Knudsen layer, while the temperature deviations $\tau_1^{(T)}$ and $\tau_2^{(T)}$ are always positive. The profiles of $\tau_1^{(T)}$ and $\tau_2^{(T)}$ are identical and do not depend on the molar fraction of the background gas. Meanwhile, both density deviations due to $X_T$ tend to increase as the molar fraction $C_0$ increases. 

The macroscopic characteristics of the helium-argon gas mixture corresponding to temperature, pressure and local molar fraction are calculated from the density and temperature deviations of species, see e.g. Ref. \cite{Fer02}. Here, the temperature, pressure and local mole fraction deviations from their equilibrium values are calculated, respectively, as 
\begin{equation}
\tau=\frac{T-T_0}{T_0}=[C_0\tau_1^{^(P)}+(1-C_0)\tau_2^{(P)}]
X_P + [C_0\tau_1^{(T)}+(1-C_0)\tau_2^{(T)}]X_T,
\lae{res1}
\end{equation}
\begin{equation}
\xi=\frac{p-p_0}{p_0}=\tau + \nu,
\lae{res2}
\end{equation}
\begin{equation}
\zeta=\frac{C-C_0}{C_0}=\frac{\nu_1^{(P)}X_P+\nu_1^{(T)}X_T}{1+\nu},
\lae{res3}
\end{equation}
where 
\begin{equation}
\nu=\frac{n-n_0}{n_0}=[C_0\nu_1^{^(P)}+(1-C_0)\nu_2^{(P)}]
X_P + [C_0\nu_1^{(T)}+(1-C_0)\nu_2^{(T)}]X_T.
\lae{res3a}
\end{equation}

The profiles of the deviations given in Eqs. (\ref{res1})-(\ref{res3}) at $T_0$=50 K as functions of distance $r/\delta$ are given in Figures \ref{fig7}-\ref{fig9} for $C_0$=0.1, 0.5 and 0.9. In each figure, three curves are provided for $\delta$=0.1, 1 and 10. The profiles were calculated by assuming that the difference between the spherical surface temperature and the reference temperature is equal to 0.1 K, as it was done in Section \ref{sec:JP_JT}. In this case, the thermodynamic forces $X_P$ and $X_T$ are equal to 0.042 and 0.002, respectively. 

An interesting feature observed in Figures \ref{fig7}-\ref{fig8} is the inversion of the temperature gradient. From the comparison of the temperature deviation for $\delta$=0.1 at different values of the molar fraction $C_0$, it can be seen that the temperature deviation is positive,  indicating that the temperature of the gas mixture increases as the distance $r/\delta \rightarrow 1$. However, for $\delta$=1 and 10, the qualitative behavior of the temperature deviation change with varying $C_0$. In case of $C_0$=0.1, the temperature deviation remains negative for both $\delta$=1 and 10, which means that the temperature of the mixture decreases as the distance $r/\delta \rightarrow 1$, i.e the inversion of the temperature gradient appears. In case of $C_0$=0.5, this inversion of the temperature gradient is observed only for $\delta$=1, while in case of $C_0$=0.9 no inversion occurs. In kinetic theory of gases, the negative temperature gradient phenomenon has already been reported in problems concerning evaporation and condensation of a single gas and it is due to the temperature jump in the Knudsen layer; see e.g. the brief review presented in the book by Sone \cite{Son48}. Molecular dynamics simulations have also predicted the inverted temperature gradient in a vapor between planar surfaces of its condensed phases, see e.g. Refs. \cite{Frez-inverted-temperature1,Meland-inverted-temperature1}. Experimental confirmation of the inverted temperature gradient is reported in Ref. \cite{Gatapova-invert-temperature-exp1}. The results of the present work show that the inversion of the temperature gradient also depends on the molar fraction of the background gas, as also shown in Ref. \cite{Sha129} for planar geometry. There is no dependence of such a phenomenon on the interatomic interaction potential.  

The local mole fraction deviation, $\xi$, is always negative, which means that the molar fraction of helium decreases in the Knudsen layer as $r/\delta \rightarrow 1$.  Meanwhile, the pressure deviation, $\zeta$, is always positive and tends to increase as $r/\delta \rightarrow 1$.

Figures \ref{fig7}-\ref{fig9} also show the temperature and pressure jumps at the interface. These jumps depend on both the rarefaction parameter and the molar fraction of the background gas.


\section{Conclusion}

The sublimation and deposition phenomena at the surface of a solid argon sphere in the presence of helium as a non-condensable or  background gas are investigated with basis of the numerical solution of the linearized Boltzmann kinetic equation. The McCormack model is employed for the collision term, while intermolecular interactions are accounted for using both the HS and AI potentials. The kinetic coefficients which determine the mass and energy flow rates are computed for three values of molar fraction of helium, namely $C_0 = 0.1$, 0.5, and 0.9, two equilibrium temperatures, $T_0 = 50$ K and 70 K, and some values of rarefaction parameter (free molecular to near viscous flow regime).  
The results show that the kinetic coefficients are highly sensitive to gas rarefaction, particularly the cross coefficient $\Lambda_{\mbox{\tiny{PT}}}$, which can deviate by more than 100$\%$  from its value in the free molecular regime. Notably, the sign of $\Lambda_{\mbox{\tiny{PT}}}$ can also change with variations in the rarefaction parameter. The background gas concentration has a significant influence on the behavior of the kinetic coefficients. Specifically, the coefficient $\Lambda_{\mbox{\tiny{PP}}}$ decreases with increasing molar fraction of the background gas, while $\Lambda_{\mbox{\tiny{TT}}}$ increases as $C_0$ varies from 0.1 to 0.5, and then decreases as $C_0$ increases further from 0.5 to 0.9.
The effect of the intermolecular interaction potential is less pronounced for the diagonal coefficients: the difference between the values obtained using the AI and HS potentials is around 15$\%$  for both $\Lambda_{\mbox{\tiny{PP}}}$ and $\Lambda_{\mbox{\tiny{TT}}}$, depending on the reference temperature. However, $\Lambda_{\mbox{\tiny{PT}}}$ exhibits a stronger dependence on the interaction potential, with differences of up to 40$\%$  between the two models, depending also on the temperature. The temperature itself has a negligible effect on the diagonal coefficients, while it significantly affects the non-diagonal coefficient $\Lambda_{\mbox{\tiny{PT}}}$. The kinetic coefficients are highly sensitive to the molecular mass ratio. Simulations performed for helium–argon and helium–krypton mixtures show a significant influence on this parameter, particularly on the $\Lambda_{\mbox{\tiny{TT}}}$ coefficient, with deviations reaching approximately 43$\%$  for $C_0 = 0.5$ and $\delta = 0.1$.

Assuming a small temperature difference between the surface of the solid sphere and the gas mixture far from it, the macroscopic parameters corresponding to temperature, pressure and local mole fraction are evaluated. The inverse temperature gradient phenomenon, previously reported by several authors, i.e., when the vapor temperature near the spherical surface is lower than that far from it, is also observed in the present simulations. In addition, temperature and pressure jumps at the solid–vapor interface are identified, and these jumps exhibit a strong dependence on both the gas rarefaction parameter and the molar fraction of species in the mixture. The dimension quantities corresponding to mass and energy flow rates, and radius of the sphere, are presented. The methodology of the present work can be used to study sublimation of droplets in processes such as ablation of metals. 
\appendix
\section{Form of collision term and related expressions}\label{sec:appendix}
\setcounter{table}{0}
\renewcommand{\thetable}{\arabic{table}}
\setcounter{figure}{0}
\renewcommand{\thefigure}{\arabic{figure}}

The form of the model collision integral proposed by McCormack in Ref. \cite{McC02} is provided below.
\[
\hat{L}_{\alpha \beta} h_{\alpha}^{(i)}=
-\gamma_{\alpha \beta}h_{\alpha}^{(i)}+\gamma_{\alpha \beta}\nu_{\alpha}^{(i)}+2\sqrt{\frac{m_{\alpha}}{m}}
\biggl[\gamma_{\alpha \beta}u_{\alpha}^{(i)}
-(u_{\alpha}^{(i)}-u_{\beta}^{(i)})\nu_{\alpha \beta}^{(1)}
\]
\[
-\left(q_{\alpha}^{(i)}-\frac{m_{\alpha}}{m_{\beta}}q_{\beta}^{(i)}\right)\frac{\nu_{\alpha \beta}^{(2)}}{2}\biggr]c_{\alpha r}
+\left[\gamma_{\alpha\beta}\tau_{\alpha}^{(i)}-2\frac{m^*}{m_{\beta}}(\tau_{\alpha}^{(i)}-\tau_{\beta}^{(i)})\nu_{\alpha \beta}^{(1)}\right]
\left(c_{\alpha}^2-\frac 32\right)
\]
\[
+\frac 45\sqrt{\frac{m_{\alpha}}{m}}\biggl[(\gamma_{\alpha \beta}-\nu_{\alpha \beta}^{(5)})q_{\alpha}^{(i)}
+q_{\beta}^{(i)}\nu_{\alpha \beta}^{(6)}-\frac 54 (u_{\alpha}^{(i)}-u_{\beta}^{(i)})\nu_{\alpha \beta}^{(2)}
\biggr]c_{\alpha r}\left(c_{\alpha}^2-\frac 52\right)
 \]
 \begin{equation}
 + 2\biggl[(\gamma_{\alpha \beta}-\nu_{\alpha \beta}^{(3)})\Pi_{\alpha}^{(i)}
+\Pi_{\beta}^{(i)}\nu_{\alpha \beta}^{(4)}\biggr]
\left(c_{\alpha r}^2-\frac{c_{\alpha t}^2}{2} \right),
\lae{ke6}
\end{equation}
where $\alpha$, $\beta=1,2$.

The quantities $\nu_{\alpha \beta}^{(n)}$ ($n=1..6$) are defined as
\begin{equation}
\nu_{\alpha \beta}^{(1)}=\frac {16}{3}\frac{m^*}{m_{\alpha}}n_{\beta}\Omega_{\alpha \beta}^{(11)},
\lae{ke11}
\end{equation}
\begin{equation}
\nu_{\alpha \beta}^{(2)}=\frac {64}{15}\left(\frac{m^*}{m_{\alpha}}\right)^2n_{\beta}\left(\Omega_{\alpha \beta}^{(12)}-\frac 52 \Omega_{\alpha \beta}^{(11)}\right),
\lae{ke12}
\end{equation}
\begin{equation}
\nu_{\alpha \beta}^{(3)}=\frac {16}{5}\frac{m^{*2}}{m_{\alpha}m_{\beta}}n_{\beta}\left(\frac {10}{3}\Omega_{\alpha \beta}^{(11)}+\frac{m_{\beta}}{m_{\alpha}}\Omega_{\alpha \beta}^{(22)}\right),
\lae{ke13}
\end{equation}
\begin{equation}
\nu_{\alpha \beta}^{(4)}=\frac {16}{5}\frac{m^{*2}}{m_{\alpha}m_{\beta}}n_{\beta}\left(\frac {10}{3}\Omega_{\alpha \beta}^{(11)}-\Omega_{\alpha \beta}^{(22)}\right),
\lae{ke14}
\end{equation}
\[
\nu_{\alpha \beta}^{(5)}=\frac {64}{15}\left(\frac{m^*}{m_{\alpha}}\right)^3\frac{m_{\alpha}}{m_{\beta}}n_{\beta}
\biggl[
\Omega_{\alpha \beta}^{(22)}+\left(\frac{15}{4}\frac{m_{\alpha}}{m_{\beta}}+\frac{25}{8}\frac{m_{\beta}}{m_{\alpha}}\right)\Omega_{\alpha \beta}^{(11)}
\]
\begin{equation}
\hskip4cm
-\frac 12 \frac{m_{\beta}}{m_{\alpha}}(5\Omega_{\alpha \beta})^{(12)}-\Omega_{\alpha \beta}^{(13)})
\biggr],
\lae{ke15}
\end{equation}
\[
\nu_{\alpha \beta}^{(6)}=\frac {64}{15}\left(\frac{m^*}{m_{\alpha}}\right)^3\left(\frac{m_{\alpha}}{m_{\beta}}\right)^{3/2}n_{\beta}
\biggl[
-\Omega_{\alpha \beta}^{(22)}+\frac{55}{8}
\Omega_{\alpha \beta}^{(11)}
-\frac 52 \Omega_{\alpha \beta}^{(12)}
\]
\begin{equation}
\hskip4cm
-\frac 12\Omega_{\alpha \beta}^{(13)}
\biggr],
\lae{ke16}
\end{equation}
where
\begin{equation}
m^*=\frac{m_1m_2}{m_1+m_2}
\lae{ke17}
\end{equation}
is the reduced mass of the mixture, and $\Omega_{\alpha \beta}^{(ij)}$ are called as $\Omega$-integrals which depend on the interatomic interaction potential. 

The parameters $\gamma_{\alpha \beta}$ appearing in the collision term given in Eq. (\ref{ke6}) are proportional to the collision frequency between species $\alpha$ and $\beta$ in the mixture. However, note that in (\ref{ke6}), the combinations $\gamma_1$=$\gamma_{11}+\gamma_{12}$ and $\gamma_2$=$\gamma_{21}+\gamma_{22}$ appear, so only $\gamma_1$ and $\gamma_2$ need to be defined. Since in the limit of a single gas, corresponding to $C_0\rightarrow 0$ or $C_0\rightarrow 1$, the McCormack model reduces to the model proposed by Shakhov \cite{Shk02} for a single gas, in which the parameter $\gamma$=$p_0/\mu_0$ appears, so by analogy, the parameter $\gamma_{\alpha}$ is chosen in the present work as
\begin{equation}
\gamma_{\alpha}=\frac{p_{0\alpha}}{\mu_{0\alpha}},
\lae{ke19}
\end{equation}
where $p_{0\alpha}$=$n_{0\alpha}kT_0$ and $\mu_{0\alpha}$ are the partial pressure and viscosity of the mixture. According to Refs. \cite{Fer02,Cha04}, the viscosity of the gas mixture in the first order approximation reads
\begin{equation}
\mu_0=\mu_{01}+\mu_{02},
\lae{ke20}
\end{equation}
where
\begin{equation}
\mu_{0\alpha}=p_{0\alpha}\frac{\Psi_{\beta}+\nu_{\alpha \beta}^{(4)}}{(\Psi_{\alpha}\Psi_{\beta}-\nu_{\alpha \beta}^{(4)}\nu_{\beta \alpha}^{(4)})},
\quad
\Psi_{\alpha \beta}=\nu_{\alpha \alpha}^{(3)}-\nu_{\alpha \alpha}^{(4)}
+\nu_{\alpha \beta}^{(4)},\quad \alpha \ne \beta.
\lae{ke21}
\end{equation}
Note that the partial viscosities $\mu_{0\alpha}$ depend on the interatomic interaction potential via the quantities $\nu_{\alpha \beta}^{(i)}$ given in (\ref{ke11})-(\ref{ke16}). For the HS model of interatomic interaction, the $\Omega$-integrals which appear in (\ref{ke11})-(\ref{ke16}) are calculated analytically as \cite{Fer02}
\begin{equation}
\Omega_{\alpha \beta}^{(ij)}=\frac{(j+1)!}{8}\left[1-
\frac{1+(-1)^i}{2(i+1)}\right]\left(\frac{\pi kT}{2m^*}\right)^{1/2}(d_{\alpha}+d_{\beta})^2,
\lae{ke18}
\end{equation}
where $d_{\alpha}$ is the diameter of species $\alpha$ and $m^*$ is the reduced mass of the mixture. In the kinetic equation for species $\alpha$ we need to specify only the diameter ratio, which is expressed in terms of the viscosity of single gases as 
\begin{equation}
\frac{d_2}{d_1}=\sqrt{\frac{\mu_1}{\mu_2}}\left(\frac{m_2}{m_1}\right)^{1/4}.
\lae{ke22}
\end{equation}
As is known, the gas viscosity depends on the gas temperature.  

For AI potential the $\Omega$-integrals must be calculated numerically from the quantum theory of interatomic interaction. In the present work, the values of $\Omega$-integrals given in Ref. \cite{Sha110} are used and they are provided in Table \ref{tabA0} for the mixtures helium-argon and helium-krypton at the reference temperatures $T_0$ considered here. The mixture helium-krypton is used to analyze the influence of the parameter $m_2/m_1$ on the solution of the problem. Note that some $\Omega$-integrals appearing in (\ref{ke11})-(\ref{ke16}) are not given in Table \ref{tabA0} because they are simplified in the collision integral (\ref{ke6}).

\section*{Acknowledgements}
D. Kalempa and I. Graur acknowledge FAPESP (Funda\c{c}\~ao de Amparo \`a Pesquisa do Estado de S\~ao Paulo) for the support of the research, grant 2022/10476-1. F. Sharipov acknowledges CNPq (Conselho Nacional de Desenvolvimento
Cient\'{\i}fico e Tecnol\'ogido), grant 303697/2014-8. The HPC resources used in this research were supported by the Information Technology Superintendence of the University of S\~ao Paulo.

\bibliographystyle{unsrtnat}

\biboptions{numbers,sort&compress}

\bibliography{papers}

\clearpage


\begin{table}
\centering
\caption{Values of $\Omega$-integrals for ab-initio potential.}
\begin{tabular}{ccc|c}\hline
 & \multicolumn{2}{c|}{He-Ar} & He-Kr \\ \cline{2-4}
$\Omega_{\alpha \beta}^{(mn)}\times 10^{16}$ [m$^3$/s] & $T_0$=50 K & $T_0$=70 K & $T_0$=70 K\\ \hline
$\Omega_{12}^{(11)}$ & 0.452601 & 0.479487 & 0.531066\\
$\Omega_{12}^{(12)}$ & 1.193073 & 1.296515 & 1.428516\\
$\Omega_{12}^{(13)}$ & 4.440914 & 4.885739 & 5.368907\\
$\Omega_{11}^{(22)}$ & 0.714749 & 0.804336 & 0.804336\\
$\Omega_{22}^{(22)}$ & 1.000645 & 1.042668 & 0.929516\\
$\Omega_{12}^{(22)}$ & 0.988153 & 1.051521 & 1.164656\\ \hline
\end{tabular}
\label{tabA0}
\end{table}

\clearpage
\begin{table}
\centering
\caption{Kinetic coefficients at temperature $T_0=$50 K (He-Ar).}
\begin{tabular}{clccc|ccc}\hline
\multirow{12}{*}{0.1} & & \multicolumn{3}{c|}{HS potential} & \multicolumn{3}{c}{AI potential}\\ \cline{3-8}
$C_0$ & $\delta$ & $\Lambda_{\mbox{\tiny{PP}}}$ & $\Lambda_{\mbox{\tiny{PT}}}$ & $\Lambda_{\mbox{\tiny{TT}}}$ &  $\Lambda_{\mbox{\tiny{PP}}}$ & $\Lambda_{\mbox{\tiny{PT}}}$ & $\Lambda_{\mbox{\tiny{TT}}}$\\ \hline
& 0$^{a}$   & 0.2422 & -0.1211 & 0.7150 & 0.2422 & -0.1211 & 0.7150 \\
& 0.01& 0.2428 & -0.1212 & 0.7140 & 0.2429 & -0.1212 & 0.7142\\
& 0.1 & 0.2481 & -0.1215 & 0.7042 & 0.2483 & -0.1218 & 0.7059  \\
& 0.5 & 0.2654 & -0.1160 & 0.6527 & 0.2667 & -0.1187 & 0.6601 \\
& 1   & 0.2810 & -0.1054 & 0.5856 & 0.2824 & -0.1085 & 0.5988\\ 
& 2   & 0.2998 & -0.0838 & 0.4795 & 0.3027 & -0.0891 & 0.4990 \\
& 5   & 0.3201 & -0.0440 & 0.3055 & 0.3267 & -0.0525 & 0.3275 \\
& 10  & 0.3182 & -0.0195 & 0.1947 & 0.3290 & -0.0288 & 0.2131 \\ \hline
\multirow{8}{*}{0.5}& 0$^{a}$   & 0.1046 & -0.0523 & 0.8964 & 0.1046 & -0.0523 & 0.8964 \\
& 0.01& 0.1048 & -0.0522 & 0.8948 & 0.1049 & -0.0523 & 0.8953 \\
& 0.1 & 0.1063 & -0.0512 & 0.8801 & 0.1068 & -0.0521 & 0.8846 \\
&0.5 & 0.1102 & -0.0440 & 0.8124 & 0.1122 & -0.0484 & 0.8316 \\
&1   & 0.1131 & -0.0346 & 0.7306 & 0.1155 & -0.0422 & 0.7637 \\ 
&2   & 0.1120 & -0.0186 & 0.6050 & 0.1164 & -0.0310 & 0.6519 \\
&5   & 0.0982 &  0.0040 & 0.3981 & 0.1057 & -0.0124 & 0.4478\\
&10  & 0.0781 &  0.0121 & 0.2608 & 0.0864 & -0.0028 & 0.2995\\ \hline
\multirow{8}{*}{0.9}&0$^{a}$   & 0.0123 & -0.0061 & 0.7272 & 0.0123 & -0.0061 & 0.7272 \\
&0.01& 0.0123 & -0.0061 & 0.7262 & 0.0123 & -0.0061 & 0.7272\\
&0.1 & 0.0123 & -0.0057 & 0.7166 & 0.0123 & -0.0059 & 0.7180\\
&0.5 & 0.0121 & -0.0038 & 0.6698 & 0.0122 & -0.0049 & 0.6754\\
&1   & 0.0111 & -0.0010 & 0.6111 & 0.0116 & -0.0037 & 0.6197\\ 
&2   & 0.0094 &  0.0018 & 0.5166 & 0.0102 & -0.0019 & 0.5266 \\
&5   & 0.0061 &  0.0036 & 0.3479 & 0.0070 &  0.0003 & 0.3560 \\
&10  & 0.0039 &  0.0032 & 0.2290 & 0.0046 &  0.0005 & 0.2343\\ \hline
\multicolumn{8}{l}{\footnotesize{$^a$ Eqs. (\ref{fm14})-(\ref{fm16}).}}
\end{tabular}
\label{tabA}
\end{table}

\begin{table}
\centering
\caption{Kinetic coefficients at $T_0=$70 K (He-Ar).}
\begin{tabular}{ccccc|ccc}\hline
\multirow{12}{*}{0.1} & & \multicolumn{3}{c|}{HS potential} & \multicolumn{3}{c}{AI potential} \\ \cline{3-8}
$C_0$ & $\delta$ & $\Lambda_{\mbox{\tiny{PP}}}$ & $\Lambda_{\mbox{\tiny{PT}}}$ & $\Lambda_{\mbox{\tiny{TT}}}$ &
$\Lambda_{\mbox{\tiny{PP}}}$ & $\Lambda_{\mbox{\tiny{PT}}}$ & $\Lambda_{\mbox{\tiny{TT}}}$ \\ \hline
& 0$^{a}$   & 0.2422 & -0.1211 & 0.7150 & 0.2422 & -0.1211 & 0.7150 \\
&0.01& 0.2428 & -0.1212 & 0.7140 & 0.2428 & -0.1212 & 0.7142 \\
&0.1 & 0.2482 & -0.1215 & 0.7041 & 0.2483 & -0.1217 & 0.7058  \\
&0.5 & 0.2658 &-0.1166 & 0.6523 & 0.2665 &-0.1179 & 0.6596 \\
&1   & 0.2809 & -0.1054 & 0.5849 & 0.2823 & -0.1081 & 0.5979 \\
&2   & 0.2996 & -0.0836 & 0.4785 & 0.3025 & -0.0884 & 0.4976 \\
&5   & 0.3196 & -0.0439 & 0.3044 & 0.3263 & -0.0511 & 0.3258  \\
&10  & 0.3172 & -0.0193 & 0.1938 & 0.3285 & -0.0273 & 0.2112 \\ \hline
\multirow{8}{*}{0.5}& 0$^{a}$   & 0.1046 & -0.0523 & 0.8964 & 0.1046 & -0.0523 & 0.8964 \\
&0.01& 0.1048 & -0.0522 & 0.8948 & 0.1048 & -0.0523 & 0.8952 \\
&0.1 & 0.1065 &-0.0512 & 0.8799 & 0.1067 & -0.0520 & 0.8842 \\
&0.5 & 0.1110 & -0.0444 & 0.8115 & 0.1121 &-0.0479 & 0.8302\\
&1   & 0.1129 & -0.0345 & 0.7290 & 0.1153 & -0.0412 & 0.7610 \\
&2   & 0.1117 & -0.0186 & 0.6025 & 0.1162 & -0.0293 & 0.6477 \\
&5   & 0.0979 &  0.0039 & 0.3952 & 0.1056 & -0.0098 & 0.4429 \\
&10  & 0.0777 &  0.0119 & 0.2583 & 0.0864 & -0.0002 & 0.2954 \\ \hline
\multirow{8}{*}{0.9} &0$^{a}$   & 0.0123 & -0.0061 & 0.7272 & 0.0123 & -0.0061 & 0.7272 \\
&0.01& 0.0123 & -0.0061 & 0.7262 & 0.0123 & -0.0061 & 0.7264 \\
&0.1 & 0.0124 &-0.0056 & 0.7164 & 0.0124 & -0.0059 &  0.7178 \\
&0.5 & 0.0120 & -0.0034 & 0.6698 & 0.0122 & -0.0048 & 0.6749\\
&1   & 0.0112 & -0.0011 & 0.6110 & 0.0116 & -0.0034 & 0.6185\\
&2   & 0.0094 &  0.0016 & 0.5161 & 0.0103 & -0.0015 & 0.5249 \\
&5   & 0.0062 &  0.0036 & 0.3472 & 0.0072 & 0.0005 & 0.3540 \\
&10  & 0.0040 &  0.0032 & 0.2283 & 0.0048 & 0.0009 & 0.2328 \\ \hline
\multicolumn{8}{l}{\footnotesize{$^a$ Eqs. (\ref{fm14})-(\ref{fm16}).}}
\end{tabular}
\label{tabD}
\end{table}

\clearpage

\begin{table}
\centering
\caption{Kinetic coefficients at temperature $T_0=$70 K, AI potential (He-Kr).}
\begin{tabular}{ccccc}\hline
$\delta$ & $C_0$ & $\Lambda_{\mbox{\tiny{PP}}}$ & $\Lambda_{\mbox{\tiny{PT}}}$ & $\Lambda_{\mbox{\tiny{TT}}}$ \\ \hline
0.1 & 0.1 & 0.2475 & -0.1214 & 0.7792 \\
    & 0.5 & 0.1042 & -0.0508 & 1.1501  \\
    & 0.9 & 0.0107 & -0.0051 & 0.8920 \\ \hline
1   & 0.1 & 0.2821 & -0.1086 & 0.6665  \\
    & 0.5 & 0.1145 & -0.0420 & 1.0091  \\
    & 0.9 & 0.0105 & -0.0032 & 0.7808 \\ \hline
10  & 0.1 & 0.3004 & -0.0292 & 0.2462  \\
    & 0.5 & 0.1017 & -0.0013 & 0.4221  \\
    & 0.9 & 0.0054 &  0.0011 & 0.3120 \\   \hline
\end{tabular}
\label{tabFkr}
\end{table}

\clearpage

\begin{table}
\centering
\caption{Flow rate $J_P$ and heat flux $J_T$ at $T_0=$50 K as function of the molar fraction $C_0$ and rarefaction parameter $\delta$ in case of $X_T$=0.002 and $X_P$=0.042 (He-Ar).}
\begin{tabular}{cccc|ccc}\hline
 & \multicolumn{3}{c|}{$J_P \times 10^2$} & \multicolumn{3}{c}{$J_T \times 10^{3}$} \\ \cline{2-7}
$\delta$ & $C_0$=0.1 & 0.5 & 0.9 & 0.1 & 0.5 & 0.9 \\ \hline
 0   & 0.9930 & 0.4289 & 0.0504 &-3.6562 &-0.4038 &1.1982 \\ 
 0.01& 0.9959 & 0.4301 & 0.0504 &-3.6620 &-0.4060 &1.1982 \\
 0.1 & 1.0185 & 0.4381 & 0.0505 &-3.7038 &-0.4190 &1.1882\\
 0.5 & 1.0964 & 0.4616 & 0.0503 &-3.6652 &-0.3696 &1.1450\\
 1   & 1.1644 & 0.4767 & 0.0480 &-3.3594 &-0.2450 &1.0840 \\
 2   & 1.2535 & 0.4827 & 0.0425 &-2.7442 & 0.0018 &0.9734\\
 5   & 1.3616 & 0.4415 & 0.0295 &-1.5500 & 0.3748 &0.7246\\
 10  & 1.3760 & 0.3623 & 0.0194 &-0.7834 & 0.4814 &0.4896 \\ \hline
\end{tabular}
\label{tabJP1}
\end{table}

\clearpage

\begin{table}
\centering
\caption{Flow rate $J_P$ and heat flux $J_T$ at $T_0=$70 K as function of the molar fraction $C_0$ and rarefaction parameter $\delta$ in case of $X_T$=0.0014 and $X_P$=0.025 (He-Ar).}
\begin{tabular}{cccc|ccc}\hline
 & \multicolumn{3}{c|}{$J_P \times 10^2$} & \multicolumn{3}{c}{$J_T \times 10^3$} \\ \cline{2-7}
$\delta$ & $C_0$=0.1 & 0.5 & 0.9 & 0.1 & 0.5 & 0.9 \\ \hline
 0   & 0.5885 & 0.2542 & 0.0299 & -2.0265 &-0.0525 &0.8656 \\ 
 0.01& 0.5900 & 0.2547 & 0.0299 & -2.0301 &-0.0542 &0.8645  \\
 0.1 & 0.6037 & 0.2595 & 0.0302 & -2.0544 &-0.0621 &0.8574  \\
 0.5 & 0.6497 & 0.2735 & 0.0298 & -2.0241 &-0.0352 &0.8249  \\
 1   & 0.6906 & 0.2825 & 0.0285 & -1.8654 & 0.0354 &0.7809  \\
 2   & 0.7439 & 0.2864 & 0.0255 & -1.5134 & 0.1743 &0.6974 \\
 5   & 0.8086 & 0.2626 & 0.0181 & -0.8214 & 0.3751 &0.5081 \\
 10  & 0.8174 & 0.2160 & 0.0121 & -0.3868 & 0.4086 &0.3484 \\ \hline
\end{tabular}
\label{tabJP2}
\end{table}

\clearpage
\begin{table}
\centering
\caption{Dimension mass and energy flow rates, and radius of the sphere at $T_0=$50 K, AI potential (He-Ar).}
\begin{tabular}{ccccc}\hline
$\delta$ & $C_0$ & $\dot{M}\times 10^{10}$ & $\dot{E}\times 10^7$ & $R_0$  \\ 
          &       & (kg/s) & (J/s) & ($\mu$m) \\ \hline
0.1 & 0.1 & 0.0019 & 0.0422 & 1.5053\\
    & 0.5 & 0.0014 & 0.0340 & 1.2758\\
    & 0.9 & 0.0002 & 0.0106 & 0.5079\\ \hline
1   & 0.1 & 0.2170 & 4.9945 & 15.053\\
    & 0.5 & 0.1478 & 3.7648 & 12.758\\
    & 0.9 & 0.0200 & 0.9929 & 5.0795 \\ \hline
10  & 0.1 & 25.647 & 652.04& 150.53\\
    & 0.5 & 11.231 & 307.70& 127.58\\
    & 0.9 & 0.8114 & 42.398& 50.795 \\ \hline
\end{tabular}
\label{tabdim}
\end{table}
\clearpage


\begin{figure}
\includegraphics[scale=1]{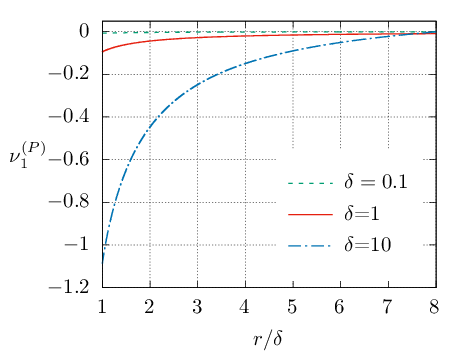}
\includegraphics[scale=1]{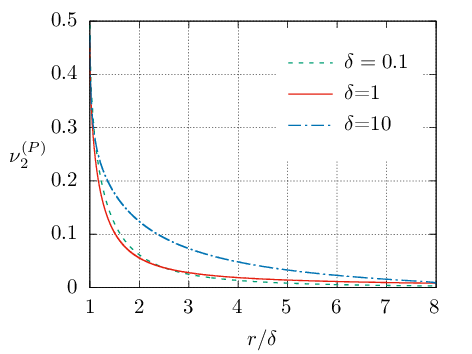}
\includegraphics[scale=1]{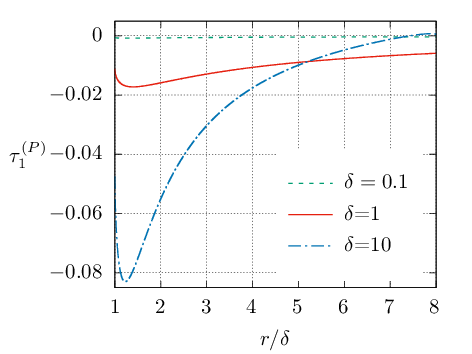}
\includegraphics[scale=1]{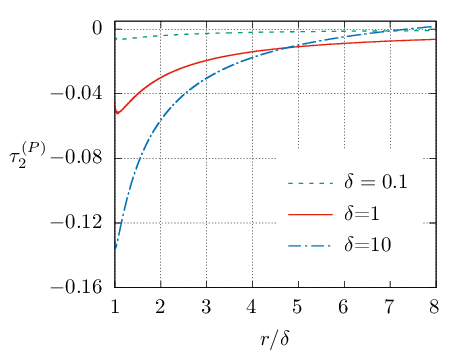}
\caption{Density and temperature deviations of species due to $X_P$ at $T_0$=50 K and $C_0$=0.1 (AI potential)}
\label{fig1}
\end{figure}

\begin{figure}
\includegraphics[scale=1]{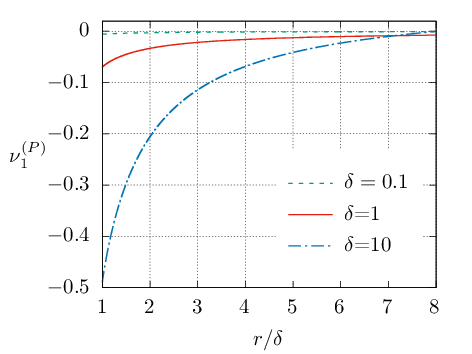}
\includegraphics[scale=1]{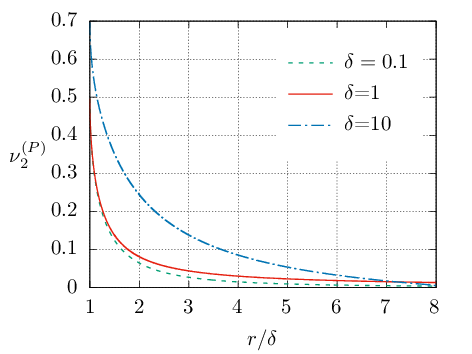}
\includegraphics[scale=1]{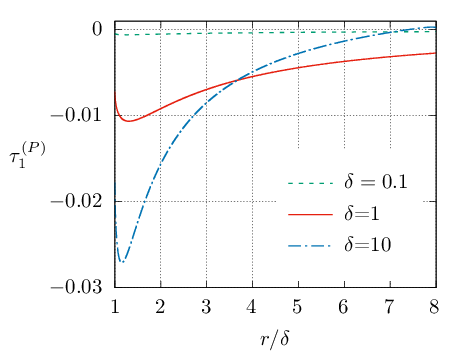}
\includegraphics[scale=1]{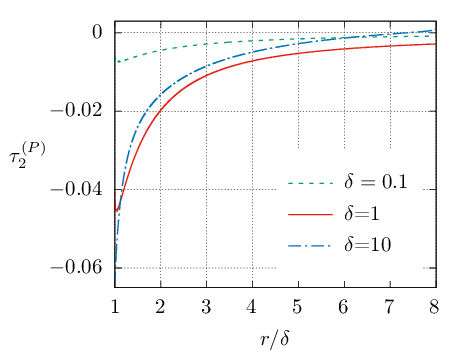}
\caption{Density and temperature deviations of species due to $X_P$ at $T_0$=50 K and $C_0$=0.5 (AI potential)}
\label{fig2}
\end{figure}

\begin{figure}
\includegraphics[scale=1]{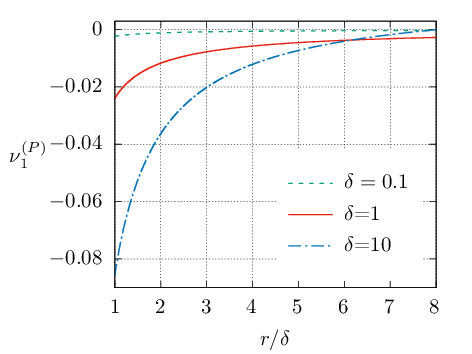}
\includegraphics[scale=1]{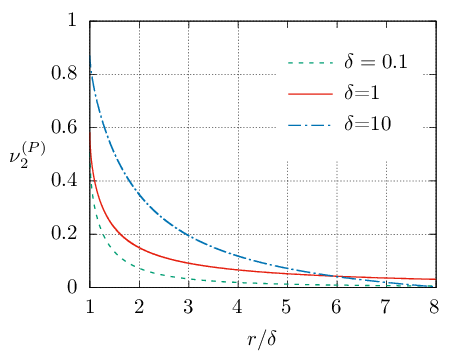}
\includegraphics[scale=1]{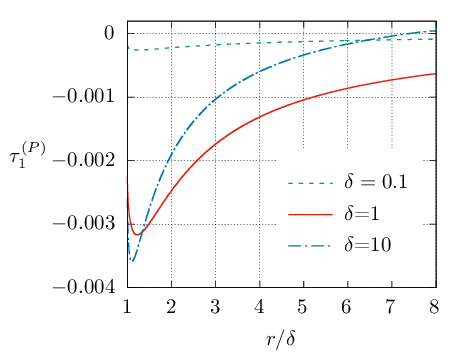}
\includegraphics[scale=1]{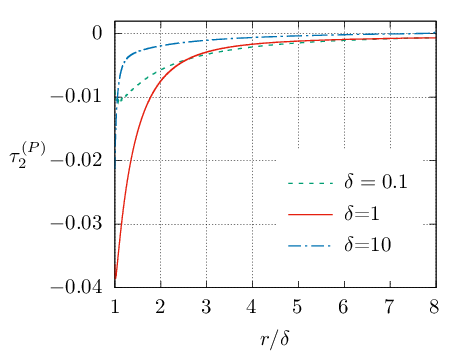}
\caption{Density and temperature deviations of species due to $X_P$ at $T_0$=50 K and $C_0$=0.9 (AI potential)}
\label{fig3}
\end{figure}

\begin{figure}
\includegraphics[scale=1]{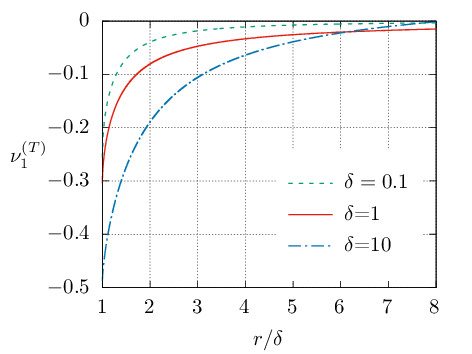}
\includegraphics[scale=1]{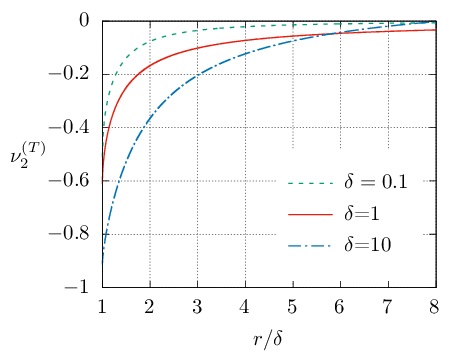}
\includegraphics[scale=1]{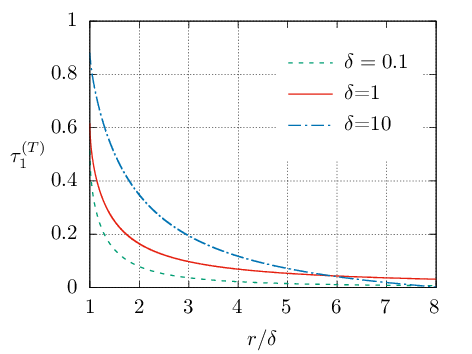}
\includegraphics[scale=1]{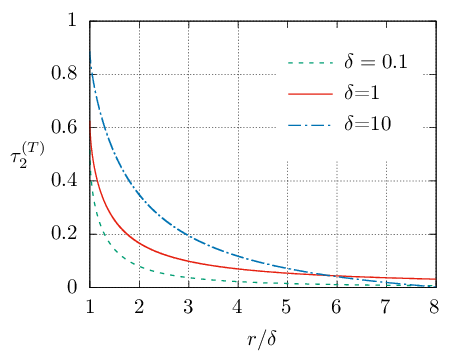}
\caption{Density and temperature deviations of species due to $X_T$ at $T_0$=50 K and $C_0$=0.1 (AI potential)}
\label{fig4}
\end{figure}

\begin{figure}
\includegraphics[scale=1]{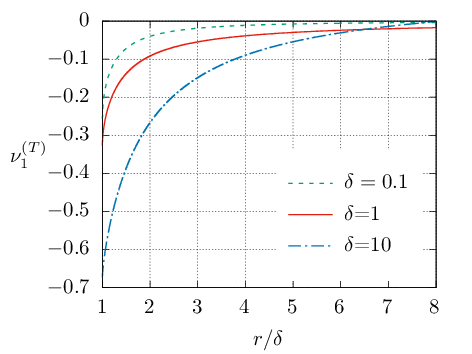}
\includegraphics[scale=1]{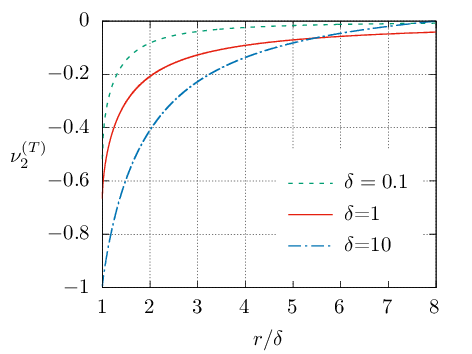}
\includegraphics[scale=1]{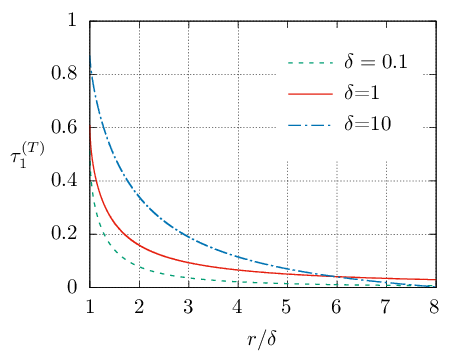}
\includegraphics[scale=1]{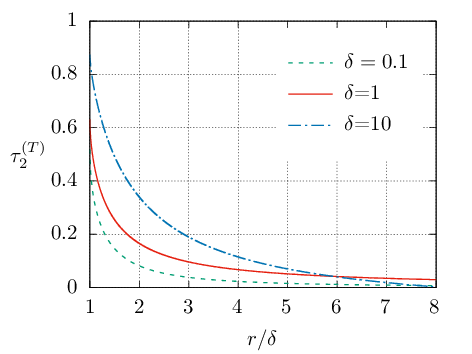}
\caption{Density and temperature deviations of species due to $X_T$ at $T_0$=50 K and $C_0$=0.5 (AI potential)}
\label{fig5}
\end{figure}

\begin{figure}
\includegraphics[scale=1]{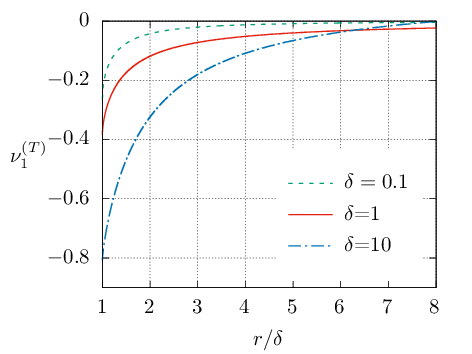}
\includegraphics[scale=1]{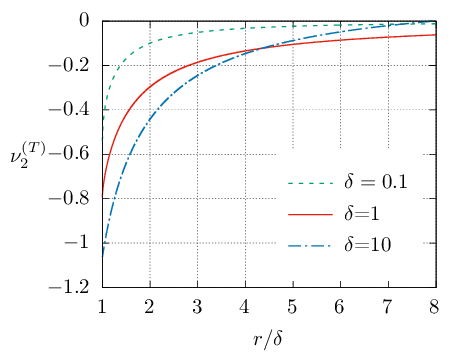}
\includegraphics[scale=1]{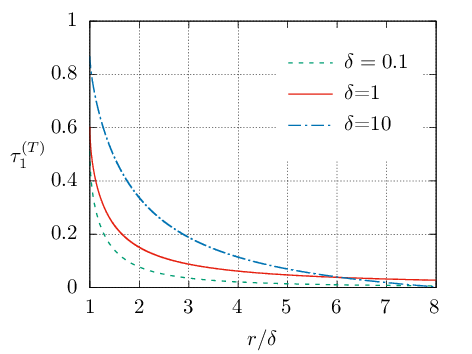}
\includegraphics[scale=1]{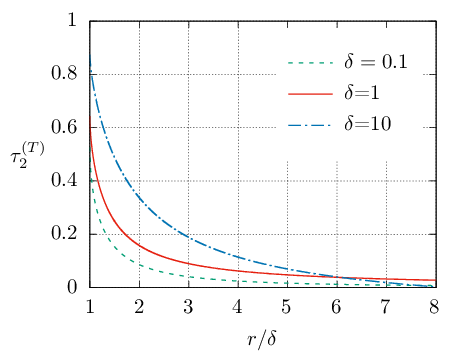}
\caption{Density and temperature deviations of species due to $X_T$ at $T_0$=50 K and $C_0$=0.9 (AI potential)}
\label{fig6}
\end{figure}


\begin{figure}
\centering
\includegraphics[scale=1.]{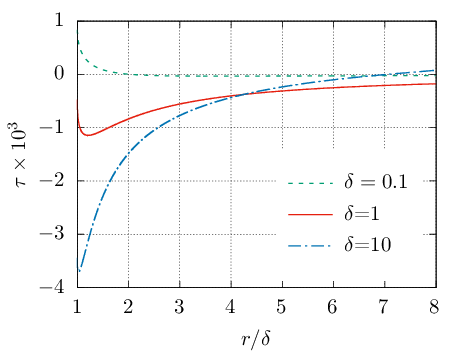}
\includegraphics[scale=1.]{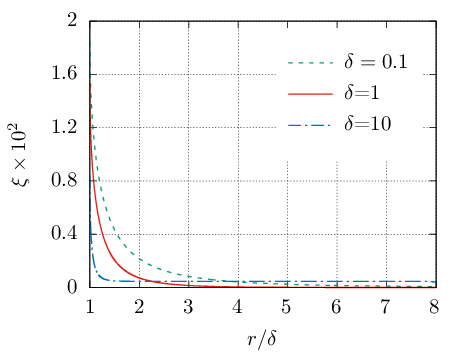}
\includegraphics[scale=1.]{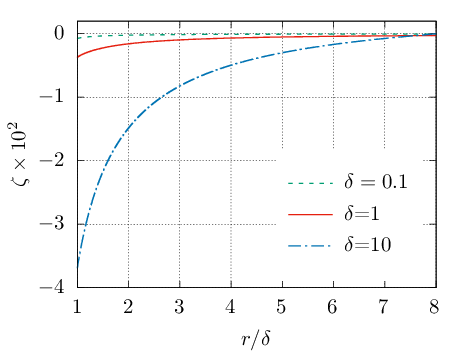}
\caption{Temperature, pressure and molar fraction deviations from equilibrium at $T_0$=50 K and $C_0$=0.1 (AI potential)}
\label{fig7}
\end{figure}

\begin{figure}
\centering
\includegraphics[scale=1.]{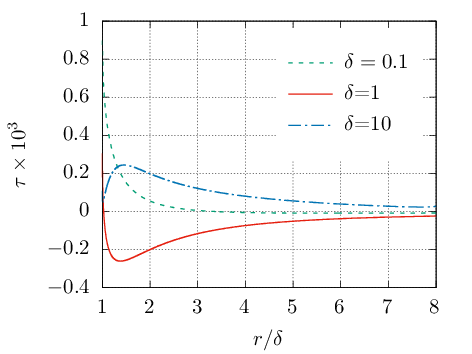}
\includegraphics[scale=1.]{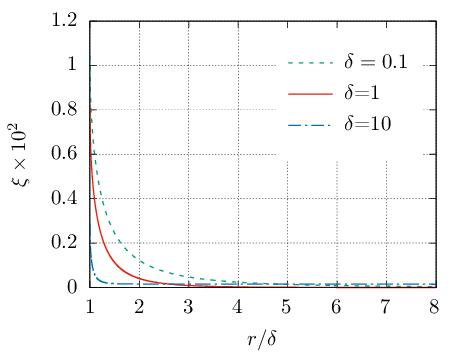}
\includegraphics[scale=1.]{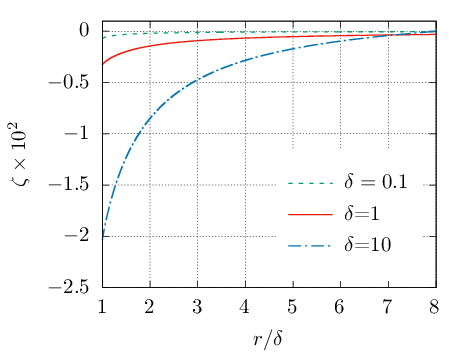}
\caption{Temperature, pressure and molar fraction deviations from equilibrium at $T_0$=50 K and $C_0$=0.5 (AI potential)}
\label{fig8}
\end{figure}

\begin{figure}
\centering
\includegraphics[scale=1.]{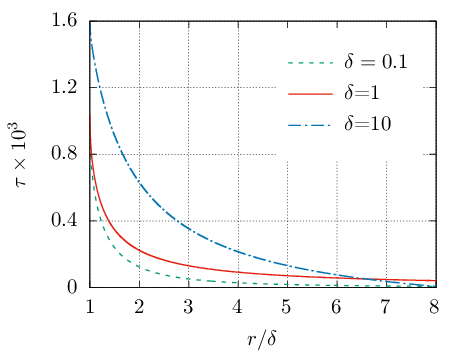}
\includegraphics[scale=1.]{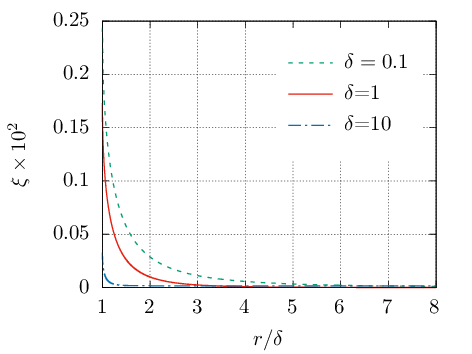}
\includegraphics[scale=1.]{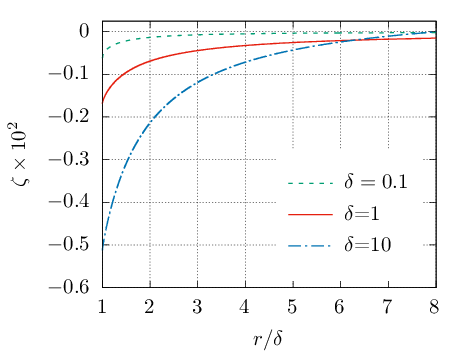}
\caption{Temperature, pressure and molar fraction deviations from equilibrium at $T_0$=50 K and $C_0$=0.9 (AI potential)}
\label{fig9}
\end{figure}

\end{document}